\documentclass[global]{svjour}[19.06.2003]
                   
\usepackage[dvips]{graphics,color}
\journalname{Astronomy \& Astrophysics Review}
\begin{document}


 
\title{Morphology and Characteristics of Radio Pulsars}

\author{John H. Seiradakis\inst{1}
\and Richard Wielebinski\inst{2}}
\institute{University of Thessaloniki, Department of Physics, Section of
Astrophysics, Astronomy \& Mechanics, GR-541 24 Thessaloniki, Greece,
\email {jhs@astro.auth.gr}
\and
Max-Planck-Institut f\"ur Radioastronomie, Auf dem H\"ugel 69,
53121 Bonn, Germany, \email {rwielebinski@mpifr-bonn.mpg.de}
}
\offprints{R. Wielebinski}          
%
\date{Received: date / Revised version: date}
\maketitle
 
\begin{abstract}
This review describes the observational properties of radio pulsars, 
fast rotating neutron stars, emitting radio waves. After the 
introduction we give a list of milestones in pulsar research. The following 
chapters concentrate on pulsar morphology:  the characteristic pulsar 
parameters such as pulse shape, pulsar spectrum, polarization and time 
dependence. We give information on the evolution of pulsars with 
frequency since this has a direct connection with the emission heights, 
as postulated in the radius to frequency mapping (RFM) concept. We 
deal successively with the properties of normal (slow) 
pulsars and of millisecond (fast - recycled) pulsars. 
The final chapters give the distribution characteristics of the presently 
catalogued 1300 objects.
\end{abstract}
   \section{Introduction}
\label{intro}
The first publication \cite{Hewish68} announcing the discovery 
of a new class of objects, soon to be known as {\it pulsars} 
\cite{Manchester77a}\cite{Lyne90a}, appeared in the literature 
almost exactly 35 years ago. Since then, large strides
toward the understanding of pulsars and their radiation have been
made. We know that they are fast rotating
%
\begin{figure}[t]
{\resizebox{11.5cm}{!}{%
\includegraphics{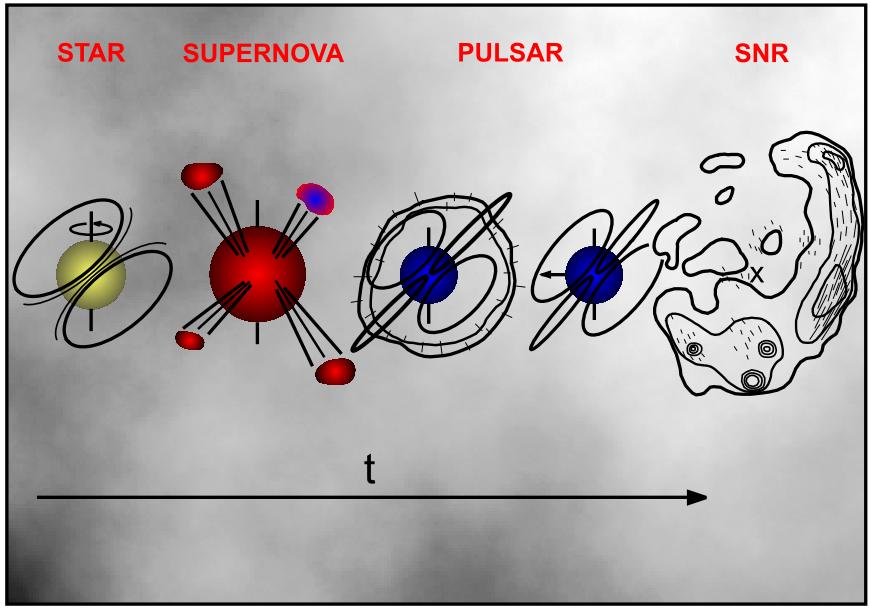}}}
\caption{\em Scenario of the birth and evolution of pulsars}
\label{Fig:1}
\end{figure}
\noindent {\it neutron stars} \cite{Baade34a}\cite{Baade34b}
\cite{Nollert88} with masses of the order of 1 \hbox{M$_{\odot}$} \cite{Pandharipande76}\cite{Datta83}\cite{Thorsett93}
\cite{Kalogera96}\cite{Callanan98}\cite{Thorsett99}
\cite{Srinivasan02}\cite{Haensel02}\cite{Jonker03}
and strong magnetic field (between $10^9$ and $10^{14}$ Gauss
\cite{Flowers77}\cite{Beskin86}\cite{Romani90}
\cite{Michel91}\cite{Arons93}\cite{Zhang00}.
They are probably born during a Supernova Type II explosion of a
(massive) late-type star (Fig. \ref{Fig:1}).\\

Pulsars emit highly accurate periodic signals (mostly in radio waves), 
beamed in a cone of radiation, centred around their magnetic axis. 
These signals reveal the period of rotation of the
neutron star, which radiates, like a light-house, once per revolution.
The lighthouse effect is caused by the dipolar magnetic field not being 
aligned with the  rotation axis of the neutron star. As a consequence of
the magnetic field, pulsar radiation is highly polarized.
Their period  of rotation (P) varies between 1.557 ms (642 Hz)
\cite{Backer82} and 8.5 s (0.12 Hz) \cite{Young99}.
As pulsars rotate, they loose energy and their rate of rotation decreases.
This period derivative ($\dot P$) is an important observational parameter.
In 1975 a pulsar in a binary system was discovered \cite{Hulse75}.
Slow rotating pulsars,
with rotation period P $\ge$ 20 ms and $\dot{P} > 10^{-18}$ are considered to be
{\it normal} pulsars. Pulsars with  P $<$ 20 ms and $\dot{P} \le 10^{-18}$ are called
{\it millisecond} or {\it recycled} pulsars. 
A full list of 1300 pulsars \cite{Manchester03} is available at 
http://www.atnf.csiro.au/research/pulsar/psrcat/. 

The period of rotation of normal pulsars increases
with time, an observational fact, discovered during the very early history
of pulsars \cite{Richards69} and led to the rejection of suggestions
that the periodic signals could be due to the orbital period of binary stars.
The orbital period of an isolated binary system decreases as it looses energy,
whereas the period of a rotating body increases as it looses energy. 
Millisecond pulsars are considered to be recycled pulsars, spun up
by mass transfer (accretion) from a binary companion \cite{Alpar82}. In the
early history of pulsars, models involving pulsations of white dwarfs and 
neutron stars were also proposed and quickly rejected.
 \begin{table}[b]
\begin{flushleft}
 
  \small
 \caption[]{Pulsar milestones}
 \begin{tabular}{|p{3.8cm}|p{5.2cm}|p{1.2cm}|}
 \hline
 Date               &Milestone                                  & Reference \\
 \hline
                      &                                                   & \\
 1932, Feb      & The discovery of neutron           & \cite{Chadwick32a}, \cite{Chadwick32b} \\
 1934, 1967    &Neutron stars are predicted        & \cite{Baade34a}, \cite{Pacini67}     \\
 1939              &Neutron stars Equation of State  & \cite{Oppenheimer39} \\
 1967, Nov 28 &The discovery  of pulsars          & \cite{Hewish68}\\
 1967, Nov 28--1968, Mar 03 &``Pulsar'' designation                 & \cite{Bell79}\\
 1968, Mar 15  &First published ``Pulsar'' designation & \cite{Telegraph68},\cite{Time68} \\
 1968                &Discovery of the Vela pulsar    & \cite{Large68a}    \\
 1968                &Discovery of the Crab pulsar    & \cite{Staelin68},\cite{Comella69}\\
 1968, Feb 24   &Dispersion measure measured   & \cite{Davies68}\\
 1968, Apr 01   &First polarimetric observation   & \cite{Lyne68a}  \\
 1968, Apr 03   &Faraday Rotation measured       & \cite{Smith68}  \\
 1968                 &Gravitational emission proposed & \cite{Weber68} \\
 1968                 &Lighthouse model proposed      & \cite{Gold68} \\
 1968, Nov         &Galactic distribution established & \cite{Large68b} \\
 1969                 &Post-detection of pulsar X-rays & \cite{Fishman69a} \\
 1969                 &Post-detection of pulsar $\gamma$-rays & \cite{Fishman69b} \\
 1969                  &Scintillation explained             & \cite{Rickett69} \\
 1969                  &Rotating Vector Model proposed& \cite{Rad69a}     \\
 1969                  &Observations of pulsar glitches& \cite{Rad69b}     \\
 1969, Aug         &First emission process proposed & \cite{Goldreich69} \\
 1974, Jul 02      &Discovery of binary pulsars      & \cite{Hulse75} \\
 1975                  &First complete theory attempted & \cite{Ruderman75}  \\
 1982                  &Discovery of millisecond pulsars  & \cite{Backer82}  \\
 1991, \~Sep 15 &Detection of extrasolar planets  & \cite{Wolszczan92}\\
 1992, Oct 19      &Detection at mm-wavelengths & \cite{Wielebinski93} \\
 1998                   &Discovery of magnetars             & \cite{Kouveliotou98} \\
 1998, Nov 05    &Discovery of the 1000$^{th}$ pulsar & \cite{csiro98} \\
                           &                                                  & \\
 \hline
 \end{tabular}
\end{flushleft}
 \end{table}

The discovery of the Vela pulsar (PSR 0833-45 -- see Table 1) led to
the suggestion of pulsar -- supernova association.
This suggestion was corroborated by the discovery of a pulsar 
(PSR 0531+21) in the heart of the Crab supernova 
remnant with a period of 33 milliseconds (see Table 1), 
which led to the unequivocal association of (radio) pulsars 
with rotating neutron stars \cite {Gold68}. Subsequent polarimetric 
observations led to the establishment of the ``Rotating Vector Model'' 
\cite {Rad69a}.
Soon after its discovery, the Crab pulsar was post-detected
in earlier (1967, Jun 04) archived X-ray data \cite{Fishman69a}
and soft $\gamma$-ray data \cite{Fishman69b}. To date (excluding
non-pulsed detections of neutron stars) 5 normal
pulsars have been detected in the optical, 17 normal 
and 6 millisecond pulsars in X-rays and 7 normal pulsars 
in $\gamma$-rays (up to $10^{25}$ Hz, covering, thus, the
largest frequency range of all known compact species emitting
intense radiation in the Universe) (updated information from 
\cite{Becker02} - W. Becker, private communication).
The first (and fastest, up to now) millisecond pulsar, PSR B1937+214
was discovered in 1982 (\cite{Backer82}. Ten years later the
first planetary system (two planets orbiting PSR B1257+12), 
outside the solar system, was discovered (\cite{Wolszczan92}).

The discovery of millisecond and binary pulsars (see Table 1)
gave new insight in pulsar research. It was soon realized
that the extremely fast rotation of millisecond pulsars could
only be explained by transfer of angular momentum from companion
stars. The notion of recycling of old and exhausted neutron stars
became popular. On the other hand binary pulsars were used to
check the effects of General Relativity, which successfully
survived the new tests. For these particular discoveries the Nobel
Price in Physics (see {\it http://www.nobel.se/physics/laureates/})
was twice awarded to pulsar researchers: In 1974
it was awarded to Antony Hewish for the discovery of the first pulsar,
PRS B1919+21, and in 1993 to Russell A. Hulse and Joseph H. Taylor 
for the discovery and subsequent work on the physics of the binary
pulsar PSR B1913+16. Recently the discovery of a double pulsar system,
PSR J0737-3039B (spin period 2.7 sec), PSR J0737-3039A (spin period 
22 msec) in a 2.4 hr eccentric orbit was announced \cite{Lyne04}. This 
system will, certainly play a very important role in deciphering the
pulsar riddle and in testing physical theories.

Meanwhile, a large number of strange and unexpected properties
of pulsar emission were observed. {\it Drifting subpulses}, {\it mode
changing}  and {\it nulling} were among the first such properties 
to be studied\cite {Rankin86a}. Extremely narrow pulses \cite{Hankins71}\cite{Kramer02}\cite{Hankins03} 
were soon to become an important tool for theoretical investigations and
$90{\hbox{$^{\circ}$}}$ polarization jumps ({\it orthogonal modes}) 
imposed restrictions to the existing models. In addition, {\it period glitches}
were observed, during which the pulsar period decreased by a large 
amount and then, within a few days, it increased again to its previous 
value. Glitch properties are used to study the physical structure of
neutron stars.
   \section{Pulsar Milestones}
Before proceeding with the description of the characteristics
of pulsar emission, it is worth looking back and paying tribute 
to the main discoveries concerning pulsars. Table 1, summarises
the most important steps in pulsar research, both in observation
and theory.

It is not always easy to unearth a ``first'' date or a ``first'' publication.
For example, J. Bell-Burnell has communicated to us her personal
view \cite{Telegraph68} and we have found a report in a 
March 1968 issue of the {\it Time} magazine \cite{Time68} referring 
to the first use of the word {\it pulsar}. Nevertheless, 
we believe that still earlier references to the ``pulsar'' notation
may have escaped our search. The list can be expanded to include
several other ``firsts''. However, we decided to restrict its size
to what we consider to be the most important ones.
   \section{Morphology}
Pulsars present to the observer a most complicated set of time variable 
phenomena. On the one hand, their period of rotation, P, varies from 
1.5 millisecond up to 8.5 seconds from object to object. On the other,
this period increases regularly with time, as the pulsar looses 
gravitational energy and slows down. The deceleration is 
%
\begin{figure}[b]
{\resizebox{0.95\textwidth}{!}{%
\includegraphics{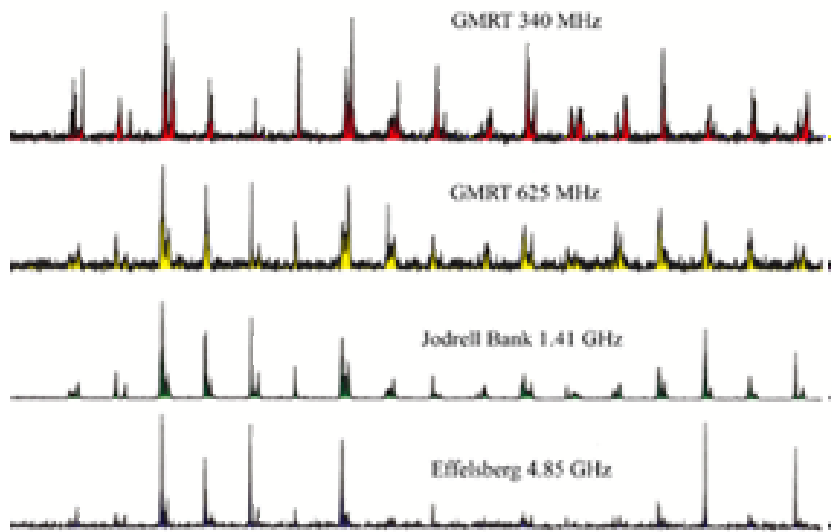}}}
\caption{\em A sequence of 17 single pulses observed simultaneously
at 4 frequencies, plotted one after the other. Courtesy of A. Karastergiou}
\label{Fig:2}
\end{figure}
expressed quantitatively by the measured $\dot P$. The youngest (normal)
pulsars exhibit usually the largest deceleration and thus they demonstrate the 
largest $\dot P$. Quite to the opposite, millisecond pulsars have very low 
$\dot P$ and are interpreted to be recycled objects, slow, normal pulsars 
that have been sped up by accretion from a binary counterpart.
The deceleration process is not always constant. Some pulsars are 
 known to exhibit glitches, sudden acceleration to higher periods, 
which are interpreted to be due to crust related ``starquakes".
 
Taking into account single pulse intensities and their duration,
extremely high brightness temperatures (of the order of $10^{29}$ K)
are calculated, especially for their low frequency emission. This
constrains their radio emission to be coherent (see also \cite{Lesch98}
and references therein). The intensity of single pulses show 
enormous intrinsic variations. The situation is even further 
complicated by the fact that interstellar scintillation introduces 
an additional fluctuating effect, in particular at lower radio frequencies. 
However with enough repeated observations the average emitted 
pulsar flux density can be determined. The addition of a sequence of 
single pulses leads quickly (usually, after a few hundred pulses) to a 
stable pulse shape, a signature of the geometry of lighthouse emission 
mechanism. The pulse shape is a firm characteristic of a pulsar, having 
anything from one up to nine sub-pulse components. 
%
\begin{figure}[b]
{\resizebox{11.5cm}{!}{%
\includegraphics{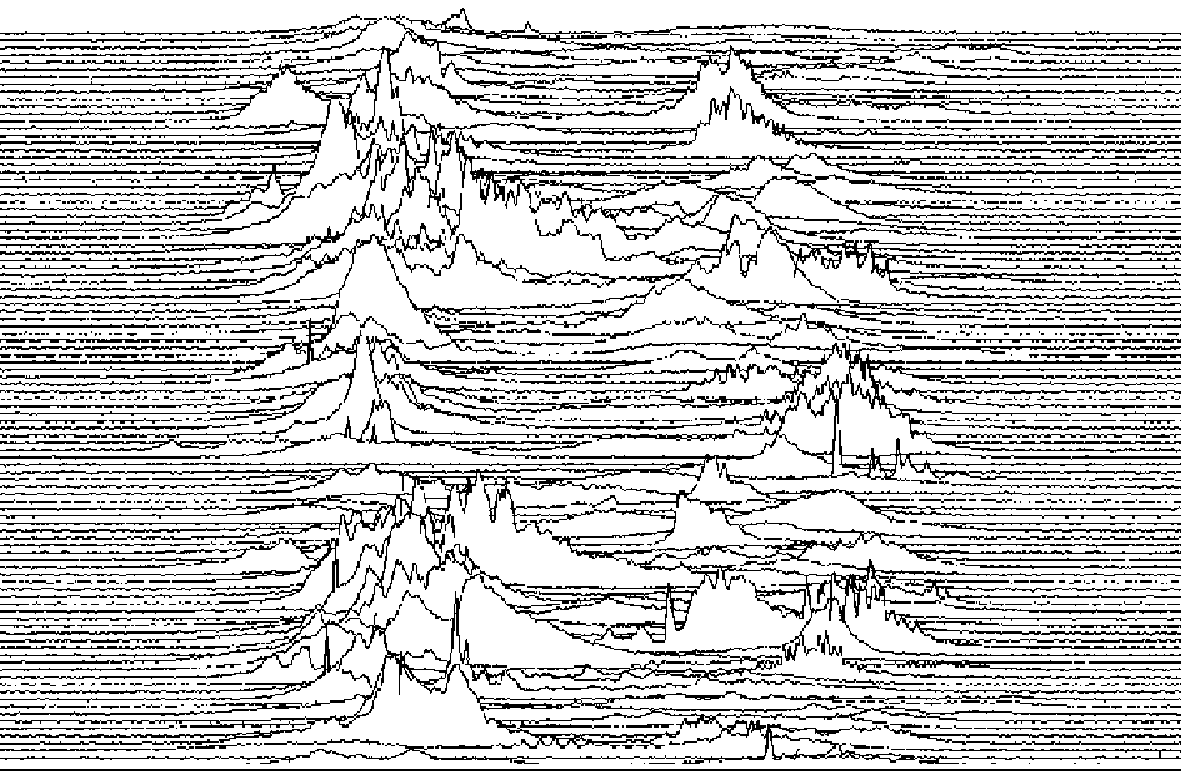}}}\\
{\resizebox{11.5cm}{3cm}{%
\includegraphics{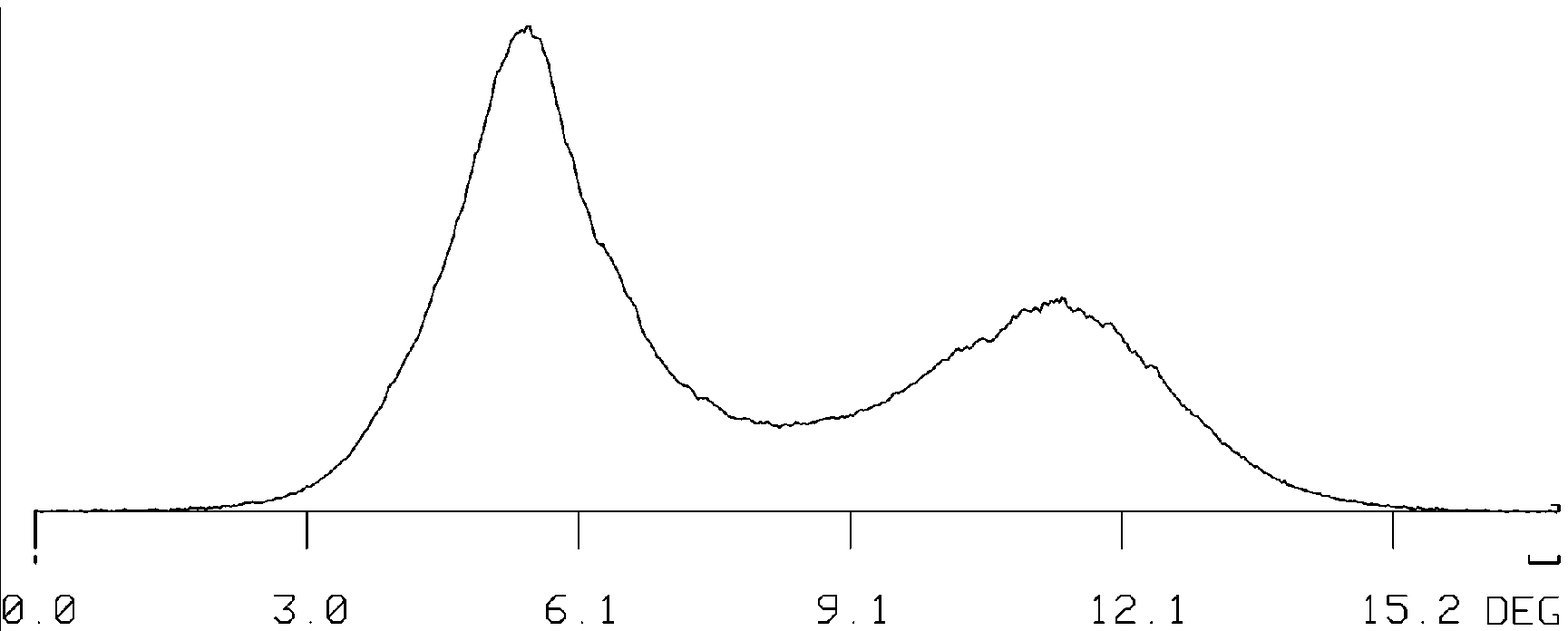}}}\\
\caption{\em (Top): A sequence of 100 single pulses from PSR 1133+16 
 plotted underneath each other (Noise wings are suppressed). 
 (Bottom): By adding the above single pulses, we get the 
 {\it\rm Integrated profile} }
\label{Fig:3}
\end{figure}
Within the sub-pulse structure a number of phenomena has been observed. 
Sub-pulse drifting (see above) reveals the emitting beam structure. In addition, 
the existence of very narrow pulsar micro-
and nano-structure \cite{Ferg78}\cite{Hankins03}, that are the signs of 
individual emission regions, has been confirmed. Pulsar radiation is highly 
polarized in a most complicated way. At low radio frequencies some pulsars 
are almost 100\% linearly polarized. Others have very high and variable 
circular polarization. The development of polarization with frequency is 
radically different from all other radio sources. The polarization may be 
high at low frequencies while dropping rapidly to zero at high frequencies. 
Possibly this is a hint for a coherent (low frequencies) -- incoherent 
(high frequencies) emission mechanism, an effect corroborated by
high frequency pulsar spectra \cite{Kramer96}.
   \subsection{Displaying pulses}
Pulsars are immediately recognised from the periodic nature of 
their radiation. Pulse sequences (Fig. \ref{Fig:2}) can be depicted in a much
more compact and informative way if their period is accurately known.
Then, instead of showing the pulses one-after-the-other, 
they can be displayed one-underneath-the-other. 
Thus, many more pulses can be conveniently accommodated 
in a single graph. By ignoring the unpulsed noise either 
side of the pulses, high time resolution single pulses are 
readily displayed (Fig. \ref{Fig:3} - top). It is evident from Figures
\ref{Fig:2} and \ref{Fig:3}
that individual, {\it single pulses} vary greatly in intensity. 
Most of these variations are intrinsic. Some are due to interstellar 
scintillation. However, if a large number of single pulses
are added together, a very stable profile is obtained 
(Fig. \ref{Fig:3} - bottom). For most pulsars, this {\it integrated profile} 
characterizes uniquely a pulsar at a particular frequency.
The stability of pulsar integrated profiles has been thoroughly
investigated for a large sample of normal pulsars \cite{Helfand75}
and some millisecond pulsars \cite{Kaspi93}.
%
\begin{figure}[b]
\rotatebox{-90}{\resizebox{7.9cm}{!}{%
\includegraphics{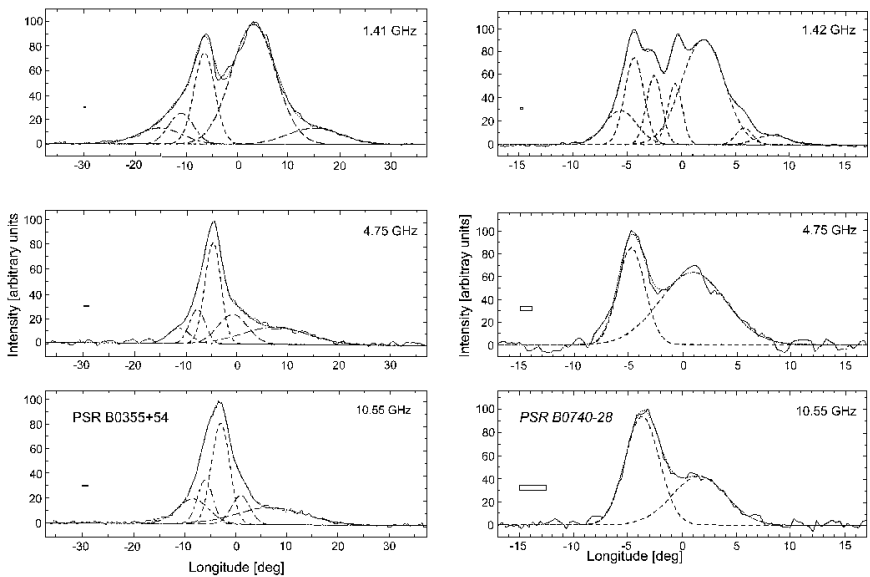}}}
\caption{\em Time-aligned profiles at three frequencies, with gauss-fitted
  components. Radius to Frequency Mapping (RFM) is obvious in this Figure.
  Left panel: PSR B0355+54. Right panel: PSR B0740-28.}
\label{Fig:4}
\end{figure}
   \subsection{Integrated Pulses}
In the integrated pulsar profiles distinct {\it components} can
be identified \cite{Seiradakis95}\cite{Kijak98}. 
They are thought to represent coherent physical regions
in the magnetosphere of the star. Therefore their properties
are of extreme importance for understanding 
the emission mechanism of pulsars.
These components are often blended and their 
shape and longitudinal location within the pulsar profile 
is difficult to establish. A large collection of pulse
profiles can be found in the European Pulsar Network
data archive at www.mpifr-bonn.mpd.de/div/pulsar/data/. 
Although there are no rigorous theoretical arguments, 
usually they can be fitted with gaussian curves, 
the parameters of which are easily
obtained. Experience has shown that pulsar profiles 
can indeed be reconstructed by a sum of individual 
gaussian components. This method usually involves some 
assumptions which can be minimized by 
reducing the number of degrees of freedom of the gauss--fitting procedure \cite{Kramer94a}\cite{Kramer94b}. 
Some gauss--fitted components are shown in Fig. \ref{Fig:4}.\\
%
\begin{figure}[h]
{\resizebox{11.5cm}{!}{%
\includegraphics{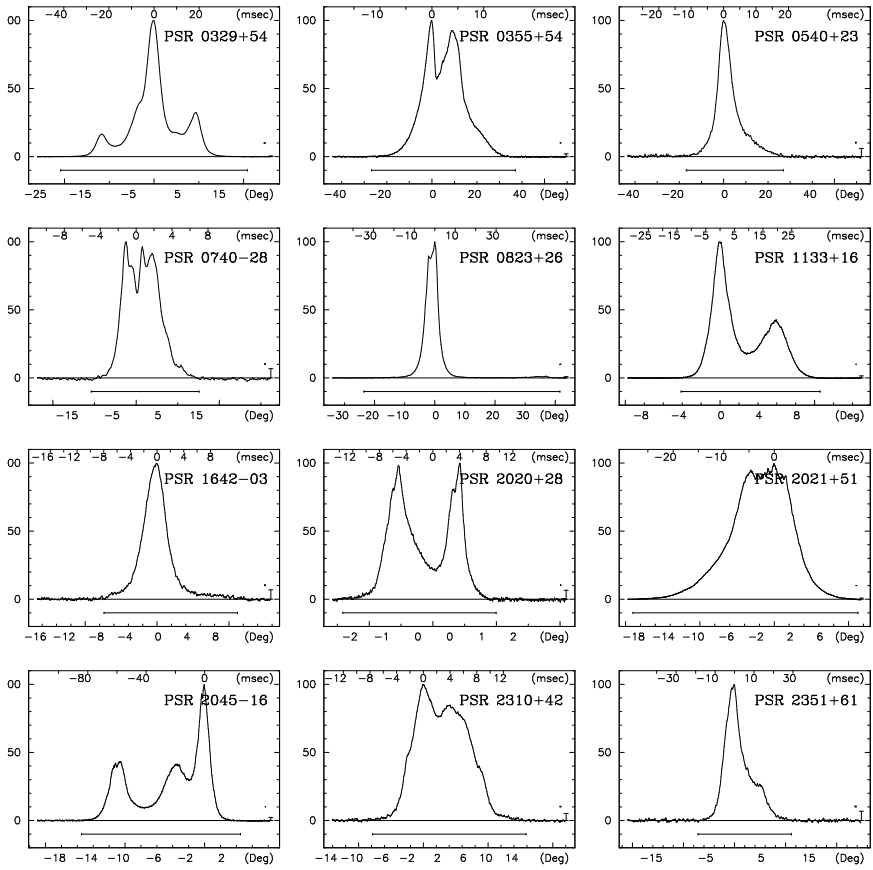}}}
\caption{\em Integrated profiles of pulsars at 1.41 GHz
 \cite{Seiradakis95}}
\label{Fig:5}
\end{figure}
%
\begin{figure}[h]
{\resizebox{11.5cm}{!}{%
\includegraphics{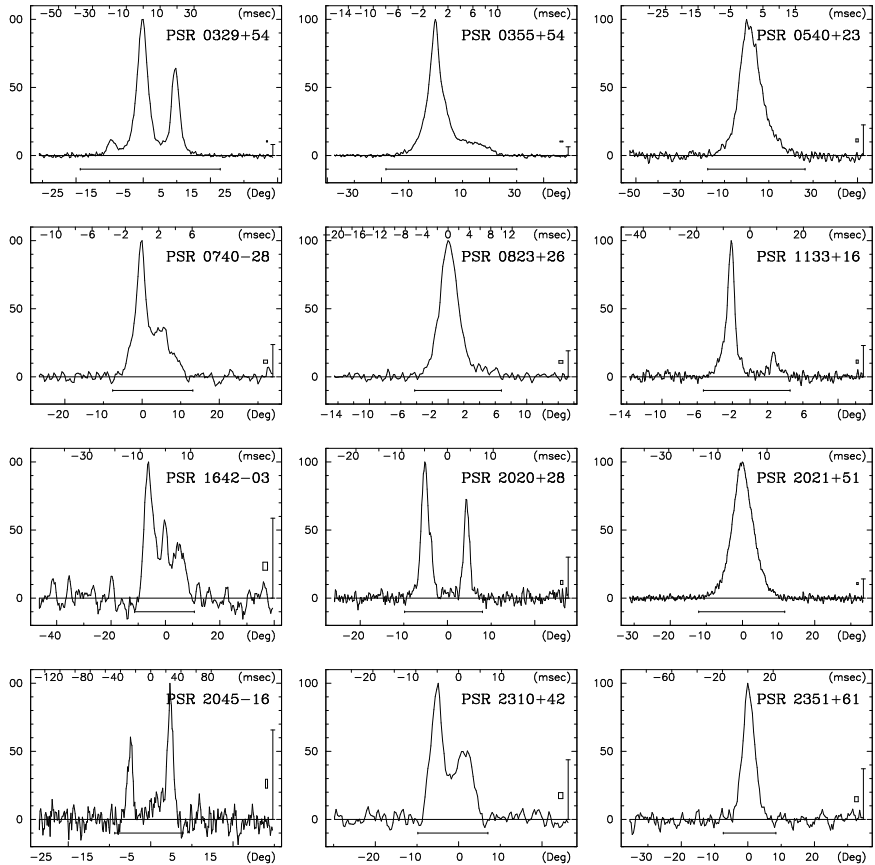}}}
\caption{\em Integrated profiles of pulsars at 10.7 GHz
 \cite{Seiradakis95}}
\label{Fig:6}
\end{figure}
   \subsubsection{Normal Pulsars (P $\ge$ 20 ms)}

Pulse profiles come in a variety of shapes (Fig. \ref{Fig:5}, \ref{Fig:6}).
In most cases they can be represented by a smooth 
curve with a single (almost gaussian) component or 
with two or more components. 
Soon after the discovery of pulsars, it was realised that their 
integrated profiles exhibit important morphological differences.
In order to explain double profiles, the {\it hollow cone
model} was proposed in the early seventies 
\cite{Komesaroff70}\cite{Backer76}\cite{Oster76}. 
Triple profiles were explained by the introduction of a central 
{\it pencil beam}  and five-component 
profiles were interpreted by assuming a more
complicated beam, comprising of a central beam 
surrounded by an {\it inner} and an {\it outer} 
cone \cite{Rankin83a}\cite{Rankin83b}\cite{Rankin93a}\cite{Rankin93b}.
The number of cones that can be accommodated within the
narrow polar cap region, whose radius is bound by the last
open magnetic field lines cannot be very large. It is interesting 
that pulsars exhibiting a single core component, seem to
occupy a distinct region in the {\it P}{$\dot P$} diagram
\cite{Gil99} (Fig. \ref{Fig:21}).
%
\begin{figure}[h]
{\resizebox{8.5cm}{!}{%
\includegraphics{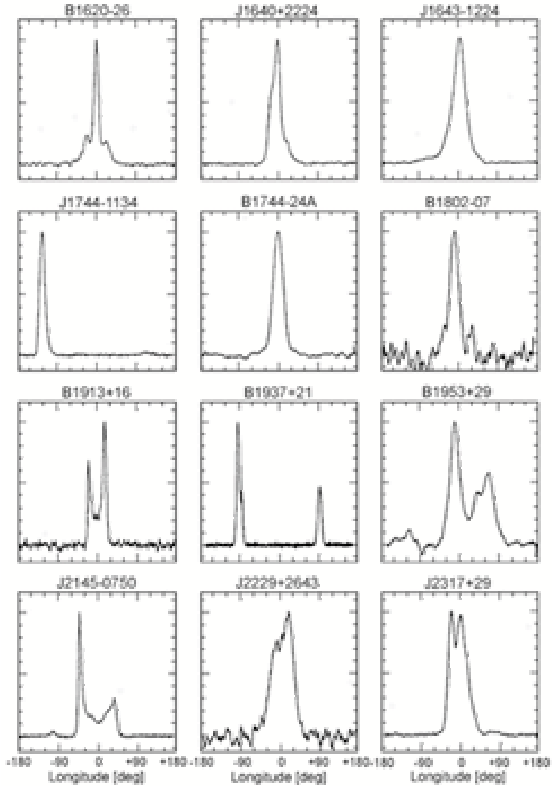}}}
\caption{\em Integrated profiles of millisecond pulsars 
 at 1.41 GHz. Effelsberg observations, M. Kramer et al.}
\label{Fig:7}
\end{figure}
On the other hand a patchy beam model was proposed
\cite{Lyne88}, according to which pulsar beams are patchy,
with components randomly located within the last open 
magnetic field lines. This model is based on the fact 
that single pulses vary in intensity and often they seem to
be missing altogether.
There have been several attempts to explain the pulsar
beam shapes using both theoretical and geometrical
arguments \cite{Gil93}\cite{Manchester95}\cite{Gil97}\cite{Mitra99}.
However, there are still many uncertainties due to the
erratic behaviour of pulsar emission and the lack of
an accepted model for pulsar radiation.
   \subsubsection{Millisecond (Recycled) Pulsars  (P $<$ 20 ms)}

The first millisecond pulsars to be discovered exhibited
rather simple profiles. Nowadays about 100 millisecond 
pulsars have been detected, many of which have complex
integrated profiles \cite{Kramer99}, not dissimilar to the
profiles of normal pulsars (Fig. \ref{Fig:7}). One difference
is that millisecond pulsars tend to have wider profiles than 
normal pulsars.
   \subsubsection{Radius to Frequency Mapping}
%
\begin{figure}[t]
{\resizebox{8cm}{!}{%
\includegraphics{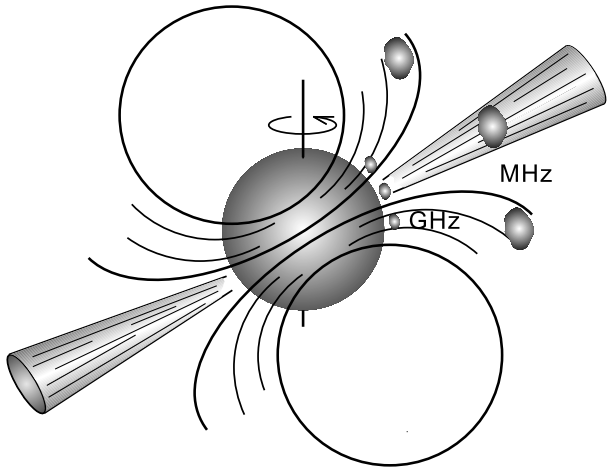}}}
\caption{\em Radius to Frequency Mapping (RFM). High frequencies
are emitted close to the surface of the star. Low frequencies are
emitted higher up, where the cone of emission is wider}
\label{Fig:8}
\end{figure}
%
\begin{figure}[h]
\rotatebox{-90}{\resizebox{9cm}{!}{%
\includegraphics{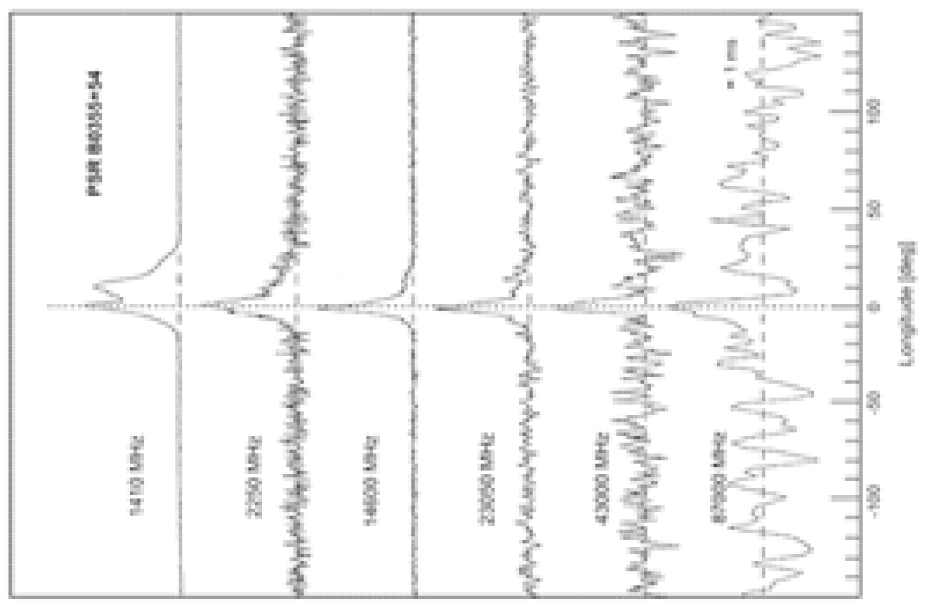}}}
\rotatebox{-90}{\resizebox{9cm}{5.5cm}{%
\includegraphics[-9,10][535,327]{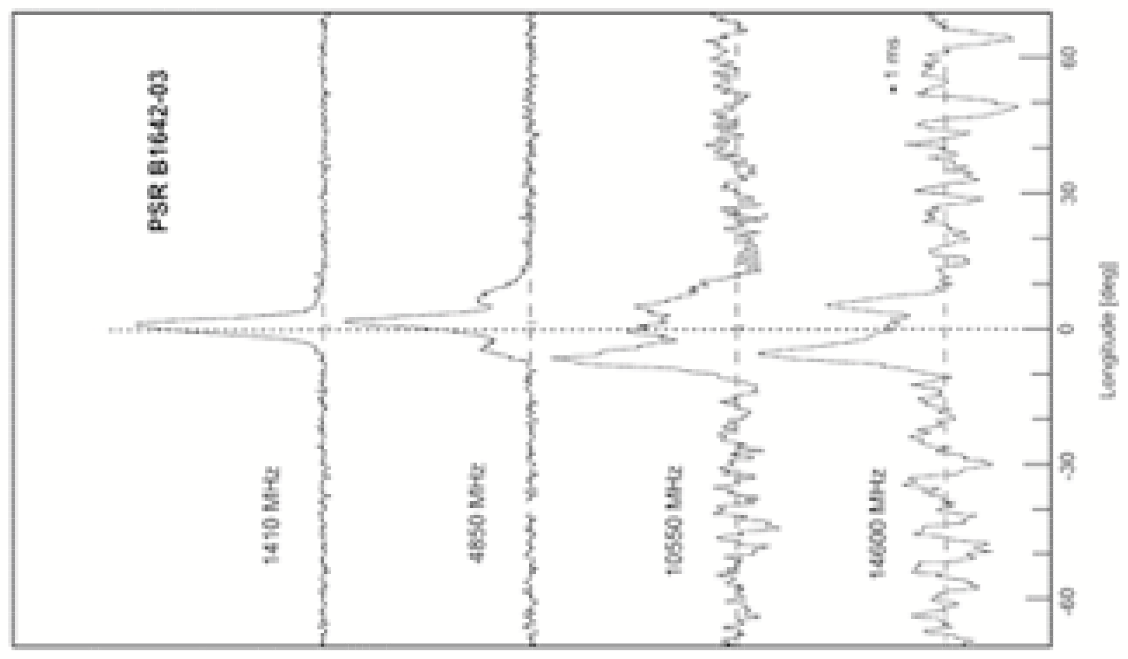}}}
\caption{\em The development of pulsar integrated profiles with 
 frequency. Note that new components can appear both at high
and low frequencies. Courtesy M. Kramer}
\label{Fig:9}
\end{figure}
\begin{figure}[h]
\rotatebox{-90}{\resizebox{8.8cm}{!}{%
\includegraphics{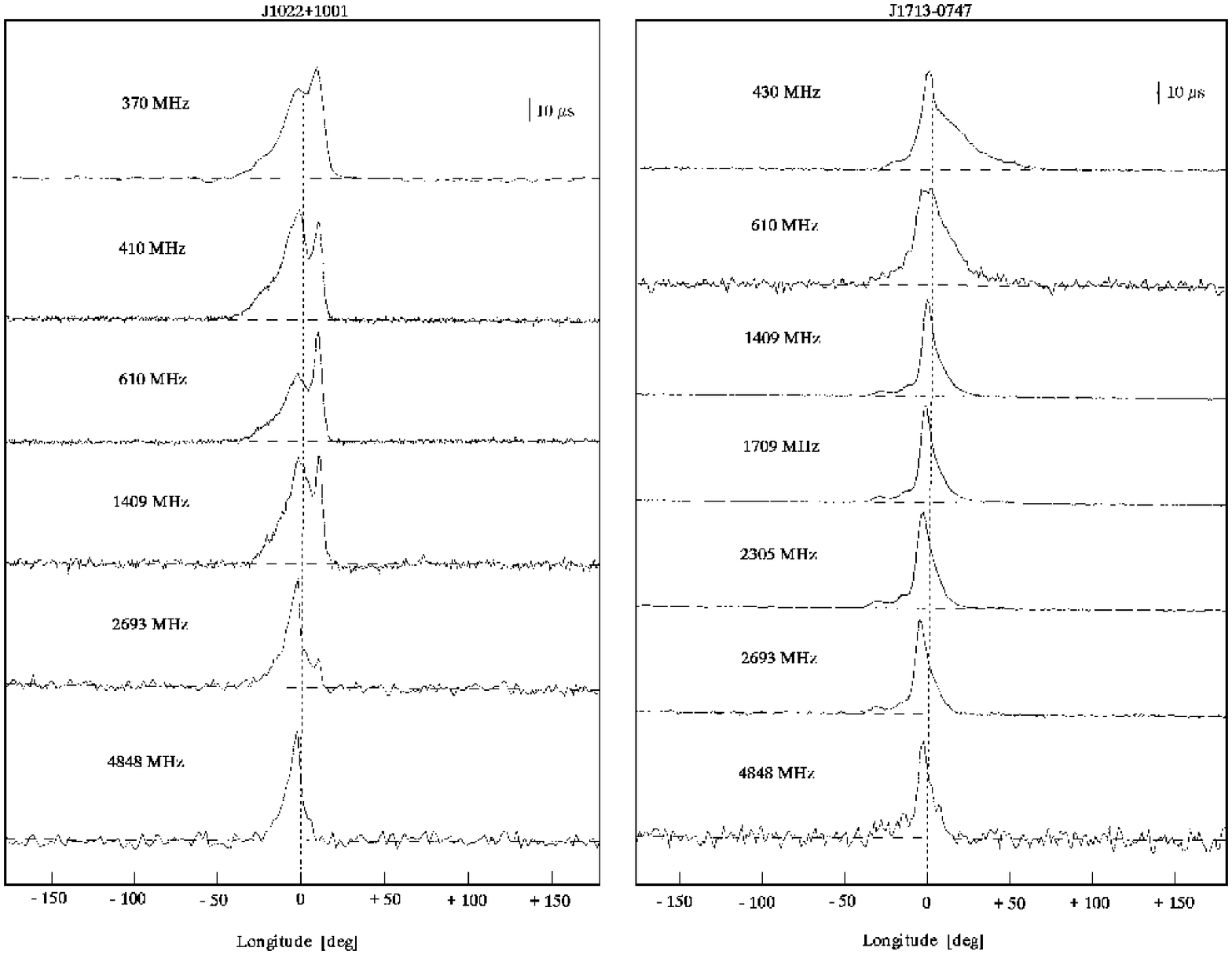}}}
\caption{\em The development of millisecond pulsars integrated 
profiles with  frequency. Adapted from \cite{Kramer99}}
\label{Fig:10}
\end{figure}

The frequency development of pulse shapes 
has led to the concept of a {\it Radius to 
Frequency Mapping} (RFM - Fig. \ref{Fig:8}), according to which, 
higher radio frequencies are emitted in the lower reaches of the pulsar 
magnetosphere (closer to the surface of the star). In order to investigate 
the RFM concept, time-aligned pulse shapes are needed, that require 
accurate pulsar timing. Following the standard pulsar model \cite{Ruderman75} 
the integrated pulse width is expected to decrease 
monotonically with frequency (RFM effect). This effect was 
implicit in earlier work \cite{Sieber75} and has been extensively
investigated ever since \cite{Cordes78}\cite{AvH97b}. Multi frequency
observations of pulsars have confirmed the narrowing of
pulse profiles with frequency \cite{Kuzmin98}\cite{Kramer99}.
The effect is adequately demonstrated for both normal
and millisecond pulsars in Figures  \ref{Fig:4}, \ref{Fig:9} and \ref{Fig:10}.
%
   \subsubsection{The Crab pulsar from radio frequencies to $\gamma$-rays}
\begin{center}
\begin{figure}[h]
{\resizebox{7.0cm}{!}{%
\includegraphics{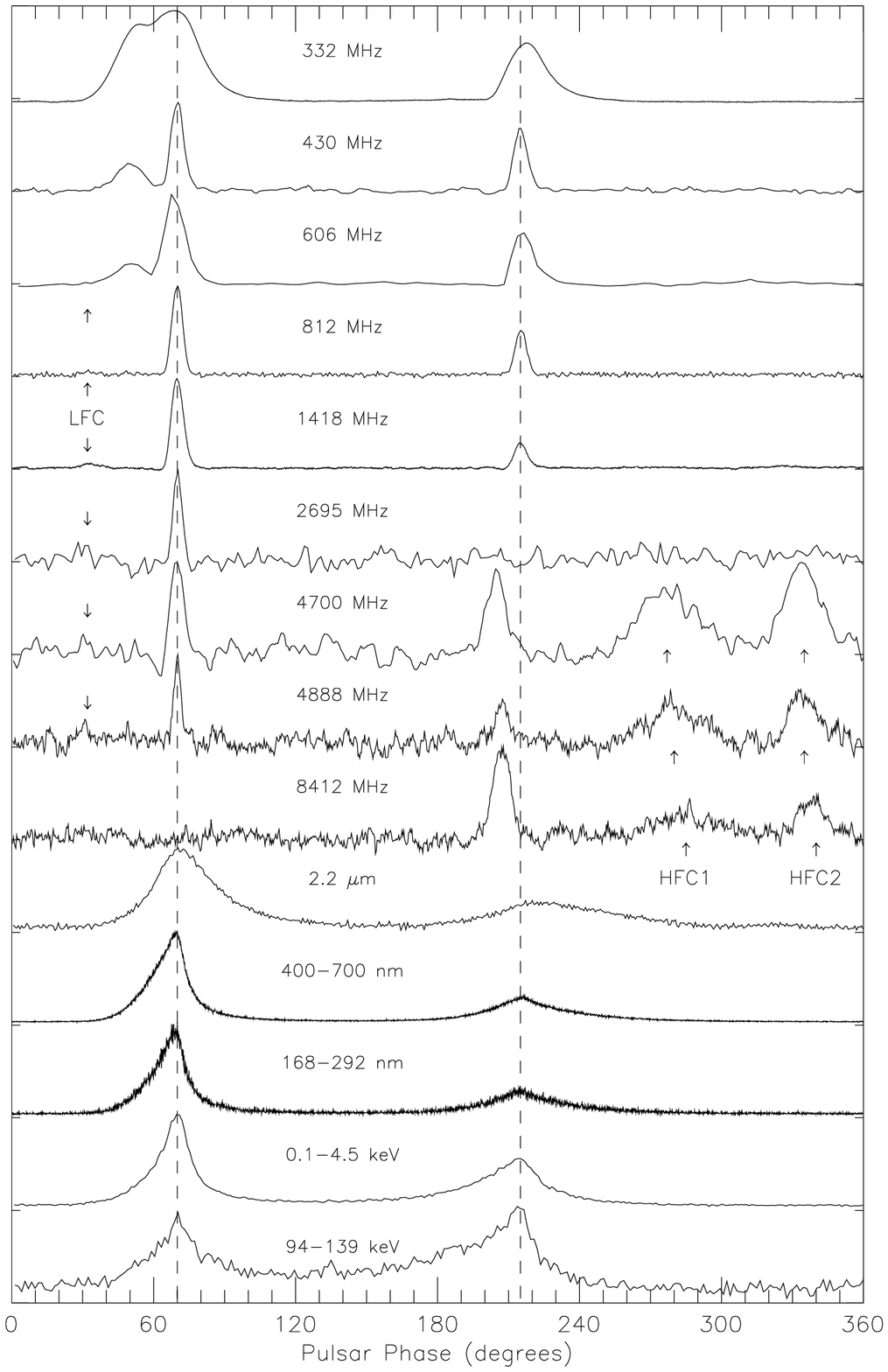}}}
\caption{\em The integrated profile of the Crab pulsar
 from low frequency radio waves to soft gamma-rays. Note the
 absence of the main pulse and the peculiar extra components 
 between 4.7 and 8.4 GHz. Courtesy T.H. Hankins}
\label{Fig:11}
\end{figure}
\end{center}

The Crab pulsar (PSR B0531+21) is probably, the best studied
pulsar. Soon after its discovery, archival searches led to post detection
of pulsed emission at X-rays and $\gamma$-rays \cite{Fishman69a},
 \cite{Fishman69b}. Its ``main-pulse -- interpulse" integrated profile
is unmistakably evident throughout the electromagnetic spectrum (Fig. \ref{Fig:11}).
However, carefully time-aligned profiles \cite{Moffett96} reveal slight,
but significant, displacement of its high frequency components from its
lower (radio) components. This has been interpreted as evidence of 
two different mechanisms of emission, with the high frequency emission
(optical, X-rays, $\gamma$-rays) originating in a region close to the
light cylinder. Furthermore, recent investigations \cite{Moffett96}
\cite{Karastergiou03c} have revealed that between 4.7 GHz and 8.4 GHz
 extra components appear in its integrated profile (Fig. \ref{Fig:11}). These
components impose additional difficulties in the investigation of the
emission mechanism of this interesting object.  
%
   \subsection{Single Pulses}
Investigations of single pulses is of utmost importance,  
showing the variability and spatial structure of  the 
pulsar emission process. Already in the discovery 
paper \cite{Hewish68} and other early papers
\cite{Davies68}\cite{Large68b}\cite{Craft68}\cite{Ekers68}
\cite{Robinson68} single pulses at low frequencies were studied. 
Single pulses show a variety of sharp 
emission structures from millisecond through microsecond down to 
nanosecond range \cite{Craft68} \cite{Hankins71}\cite{Lange98}
\cite{Hankins03}. Early two-frequency simultaneous 
observations suggested that the emission is inherently broad-band 
\cite{Bartel78a}, i.e. emission is correlated over a wide frequency 
range. Later observations \cite{Boriakoff81a}\cite{Bartel81}
\cite{Davies84}\cite{Kardashev86} were made for total 
intensity only and at most  for two simultaneous frequencies. 
Most of these investigations were then used to determine the 
bandwidth of the emission. More recently, using many radio 
telescopes at different frequencies, simultaneously, at up to five 
frequencies \cite{Kramer03} have investigated the 
single pulse characteristics in detail (see also Fig. \ref{Fig:2}).   
   \subsection{Millistructure, Microstructure, Nanostructure}
The fact that very short time structures are present 
in single pulses was noted from the very beginning \cite{Hewish68}  
leading to a stringent requirement for theories attempting 
to explain the origin of pulsar emission. First observation 
of pulsar millistructure was made soon after their
discovery \cite{Craft68}. The limitation was 
due to signal to noise in the narrow band receiver 
needed to show such short time structures. 
A few years later a de-dispersion technique at the 
frequency of 115.5 MHz revealed time structures as
short as 8 $\mu$s in PSR 0950+08 \cite{Hankins71}. 
Direct observations at 1420 MHz on PSR 1133+16
\cite{Ferg76} resolved structures with time 
scale of ~ 14 $\mu$s. The microstructure was 
found to be broad-band \cite{Rickett75}\cite{Boriakoff81a}. 
 Periodic structures were observed that seem to be 
also correlated across a wide 
frequency range \cite{Bor81b}. More recently 
\cite{Lange98} microstructure investigations were 
extended up to 4.85 GHz, showing that many pulsar have 
this emission signature. This result was confirmed in the 
studies of the Vela pulsar which shows microstructure 
\cite{Kramer02} in most of the pulses. Most recently 
the time structure studies were taken in to the 
nanosecond range with observations of giant pulses 
from the Crab pulsar \cite{Hankins03}. Their 
best time resolution was in fact 2 nanoseconds. This latest 
observation suggests that the plasma responsible 
for such emission must be of the order of one meter 
in size. If the emission is isotropic, these nanosecond pulses
must be the brightest transient source in the radio sky.
   \section{Flux densities}
The flux density of a radio source and its frequency 
evolution (its spectrum) are basic information that 
relate to the emission mechanism. However, one of 
the problems with flux densities is that pulsars vary on
various time scales: (a) due to inherent variations 
\cite{Stinebring00} or (b) due to scintillations 
\cite{Malofeev00} or (c) due to scattering \cite{Lohmer02}. 

For low frequency radio waves of the Milky Way the spectrum 
of the radio emission could be explained only by the non-thermal 
(synchrotron) emission process. From the very 
beginning of pulsar observations it was clear that 
the spectra of pulsars were very different from all 
other known radio sources \cite{Lyne68b}\cite{Robinson68}. 
Instead of values of $\alpha \sim$ -- 0.8
( S = $\nu ^{\alpha}$),  as in cosmic radio sources, the 
observed spectral index of pulsars was $\alpha \sim$ -- 1.5 
on average. The spectral index at the highest 
frequencies was for some objects much higher 
than this average value. At lower radio frequencies 
a spectral turn-over was observed. It was known 
in the earliest papers that pulsars were very weak 
at higher frequencies posing in fact a great instrumental 
challenge to study these objects at cm wavelengths. 
%
\begin{center}
\begin{figure}[h]
{\resizebox{7.5cm}{!}{%
\includegraphics{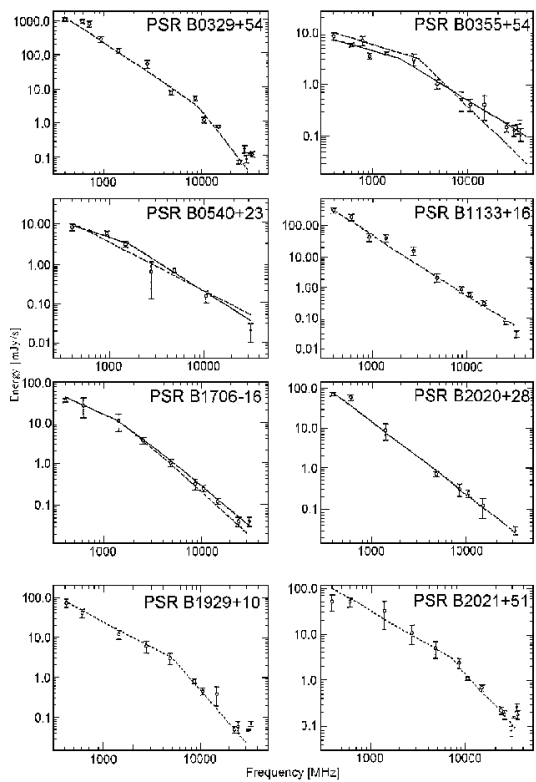}}}
\caption{\em Spectra of pulsars detected at 9 mm. \cite{Wielebinski02}}
\label{Fig:12}
\end{figure}
\end{center}
   \subsection{Normal Pulsars (P $\ge$ 20 ms)}
The first spectra of pulsars, using flux density 
values at three (plus an upper limit at 1.4 GHz) 
frequencies, were obtained in 1968 \cite{Lyne68b}. 
Earlier observations \cite{Robinson68}, published 
slightly later, were obtained at five frequencies, four 
of them simultaneously, giving an average spectrum and 
spectra of individual (single) pulses of PSR1919+21. 
The average spectrum was extended to 2.7 GHz 
and suggested a spectral break with an index of 
$\alpha \sim$ -- 3.0 above 1.4 GHz. The data 
collection to determine flux densities of a 
larger sample of pulsars took many years to complete. 
While numerous observatories (Arecibo, 
Jodrell Bank, Green Bank, Parkes) made observations 
at frequencies of 1.4 GHz and below 
only the Goldstone facility detected pulsars at 13cm 
 \cite{Ekers68}. Observations of the low frequency 
extension of pulsar spectra were carried out in the 
Soviet Union \cite{Brezgunov71}\cite{Bruck73}\cite{Malofeev00} 
at frequencies as low as 10 MHz. Three pulsars were detected 
at 8.1 GHz \cite{Huguenin71}. The suggested existence 
of a spectral break at high radio frequencies 
\cite{Robinson68} was later confirmed \cite{Backer72}.

A major step forward in the measurement of pulsar 
flux densities at high radio frequencies and 
hence of pulsar spectra was made by the 
commissioning of the 100-m Effelsberg radio 
telescope. Immediately, six pulsars were detected 
at 2.8 cm wavelength \cite{Wielebinski72}. This 
telescope continued to set records of the 
highest frequencies at which pulsars could be studied 
by reporting detections at 22.7 GHz 
\cite{Bartel77}\cite{Bartel78b} and finally in the mm-wavelengths 
\cite{Wielebinski93}\cite{Kramer97}. A detection of 
the pulsar PSR 0355+54 at 3 mm wavelength has
also been achieved  with the Pico Veleta telescope 
\cite{Morris97}. 

The early measurements of flux density 
of the strongest pulsars had to give way to studies of 
larger samples, if possible with a wide frequency 
flux density coverage. An early compendium of pulsar 
spectra was given for 27 pulsars \cite{Sieber73}.
Further multi-frequency spectra have been
presented \cite{Backer74}\cite{Sieber75}. In both of these 
papers the spectral breaks of some pulsars at high 
frequencies were confirmed and in the latter work the 
frequency evolution of the pulse width was 
noticed (this eventually led to the {\it Radius-to-Frequency 
Mapping} concept). Low radio frequency observations 
\cite{Malofeev80} confirmed the 
cut-off in pulsar spectra for a number of objects.

Subsequently, the flux density of a larger sample of 
pulsars was measured at several frequencies
\cite{Seiradakis95}\cite{Lorimer95}\cite{Kijak98} and
spectra of pulsars were derived \cite{Malofeev94}\cite{Toscano98}
\cite{Maron00} (Fig. \ref{Fig:12}). From all these publications the 
conclusion was that the average spectral index is
$< \alpha >$ = --18. From our gauss fitted
distribution (see below) we have found a similar value, 
$< \alpha >$ = --1.75. It is obvious from Figure 26, that the distribution
is fairly wide.
Some 10\% of all pulsars require a two power law fit in the high 
frequency range. A small number of pulsars have been recognized 
with almost flat spectrum ($\alpha >$ --1.0) \cite{Maron00}. 
In addition pulsar spectra seem to follow the power law
down to low frequencies (a few 10s of MHz) with a few exceptions,
where a turn-down is observed.
    \subsection{Millisecond Pulsars (P $<$ 20 ms)}
Millisecond pulsars were discovered in 1982
\cite{Backer82} as a result of  a search in the 
direction of radio sources with very steep spectra. 
The flux density of these objects is 
very low. This, combined with the effects of interstellar 
broadening, rendered their detection difficult. 
Early studies of millisecond pulsars 
\cite{Erickson85}\cite{Foster91} suggested that 
these objects have spectra steeper than 
slow pulsars. Recently high frequency
observations of millisecond pulsars were also 
made \cite{Kijak97}\cite{Maron04} in
spite of their low flux densities. The spectra of 20 objects
were studied at lower radio frequencies in a survey of 
280 pulsars \cite{Lorimer95} in the northern hemisphere. 
Several southern millisecond pulsars were also studied
 \cite{Toscano98} in a narrow frequency range. 
A major study of millisecond pulsars 
\cite{Kramer98}\cite{Xilouris98}\cite{Kramer99} 
showed that once a volume-limited sample is
considered, many of the characteristics of slow and 
fast pulsars (spectrum, pulse shapes, number of sub-pulse 
components, polarization) are the same. The one distinct 
difference is found in the luminosity.
Slow pulsars are some 10 times more luminous than 
millisecond pulsars. An investigation of 
the low frequency turn-over of millisecond pulsars 
\cite{Kuzmin01}\cite{Malofeev03} revealed that the 
morphology of millisecond pulsars is very similar to 
that of normal pulsars but with lower luminosity.
%
\begin{center}
\begin{figure}[b]
{\resizebox{8cm}{!}{%
\includegraphics{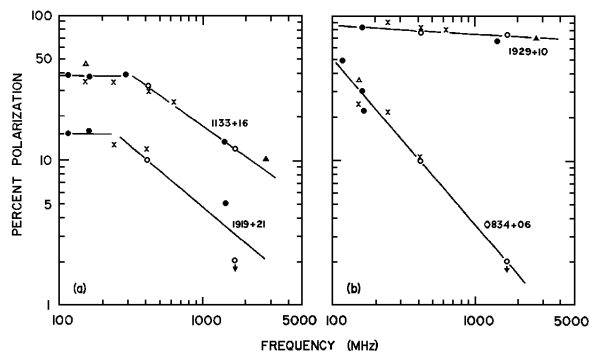}}}
\caption{\em Early pulsar polarization \cite{Manchester71a} showing
the frequency dependence}
\label{Fig:13}
\end{figure}
\end{center}
   \section{Polarisation}
%
\begin{center}
\begin{figure}[b]
{\resizebox{9.5cm}{!}{%
\includegraphics{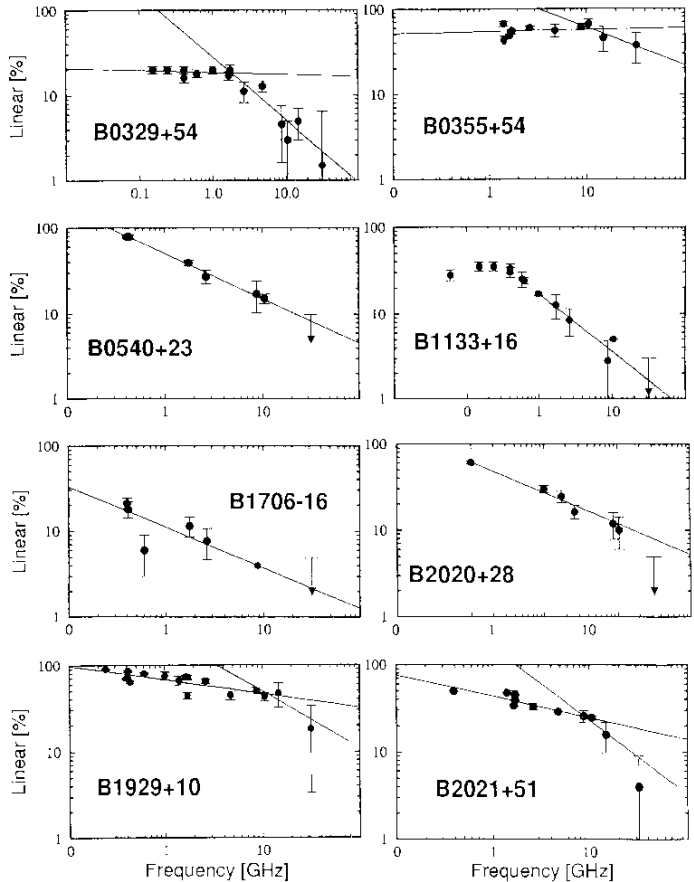}}}
\caption{\em Polarisation development with frequency of 
the integrated pulses of normal pulsars for the sample of 
pulsars shown in Fig. \ref{Fig:12} \cite{Wielebinski02}}
\label{Fig:14}
\end{figure}
\end{center}
Soon after the discovery of pulsars \cite{Hewish68}
their linear polarization was also discovered \cite{Lyne68a} 
using the Jodrell Bank Mark I radio telescope. Variations 
of the intensity from pulse to pulse were detected 
from orthogonal dipoles connected to a high speed recorder. 
The linear polarization was found to be surprisingly 
high even at lower (150 MHz; 408 MHz) radio 
frequencies. Soon it was realized that considerable 
circular polarization was also present in pulsar emission 
\cite{Craft68}\cite{Clark69}. The early observations 
with the Parkes telescope \cite{Rad69a} of the 
pulsar PSR B0833-45 showed a very high degree 
of linear polarization of the integrated pulse (in fact 
nearly 100\%) and gave arguments for a (magnetized) {\it rotating 
vector model} for pulsars. The observations that followed, 
e.g. \cite{Ekers69}\cite{Morris70}, showed that the 
phase-drift of the linear polarization is a common
feature in pulsars and, hence, gave support for 
the rotational model.
   \subsection{Integrated Pulses}
After these early observations a number of 
observers embarked on determining the detailed polarization 
characteristics of larger samples of pulsars. 
In all these studies all four Stokes parameters were 
observed since pulsars unlike extra-galactic 
sources showed high degree of linear and circular polarization. 
   \subsubsection{Normal Pulsars (P $\ge$ 20 ms)}

In 1971 the results of observations of 21 pulsars, 
at the frequencies of  410 and 1665 MHz were 
%
\begin{figure}[t]
{\resizebox{10.0cm}{!}{%
\includegraphics{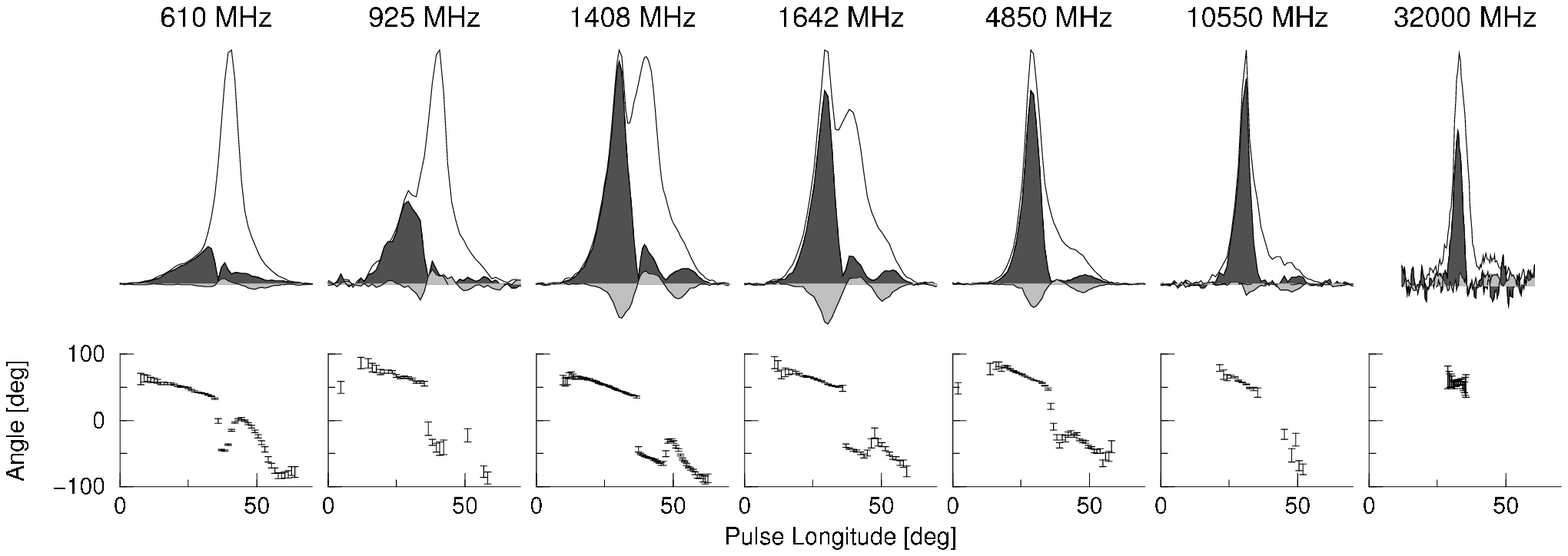}}}
{\resizebox{10.0cm}{!}{%
\includegraphics{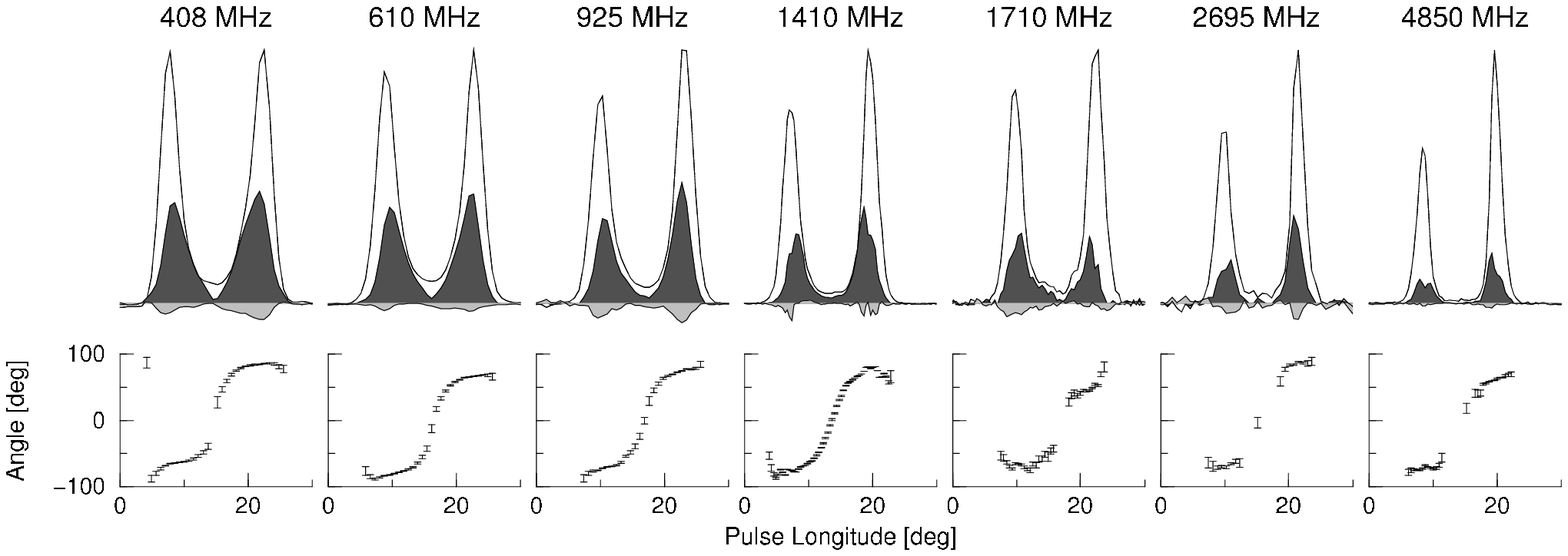}}}\\
{\resizebox{10.0cm}{!}{%
\includegraphics{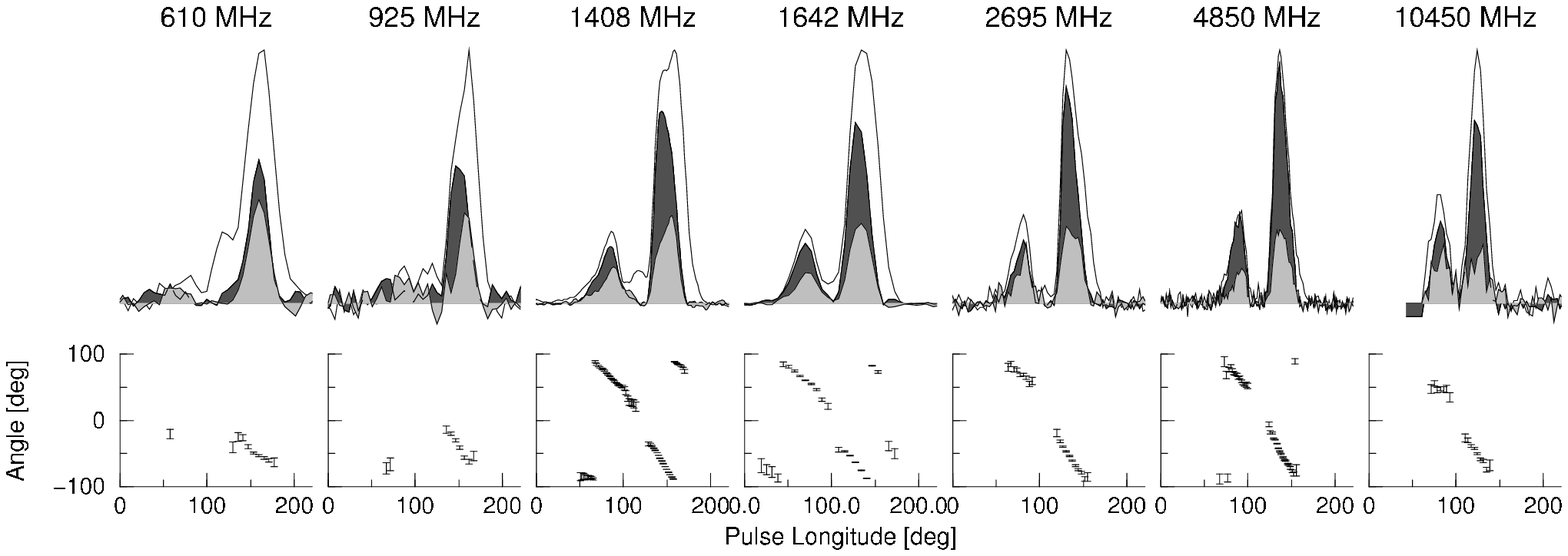}}}
{\resizebox{10.0cm}{!}{%
\includegraphics{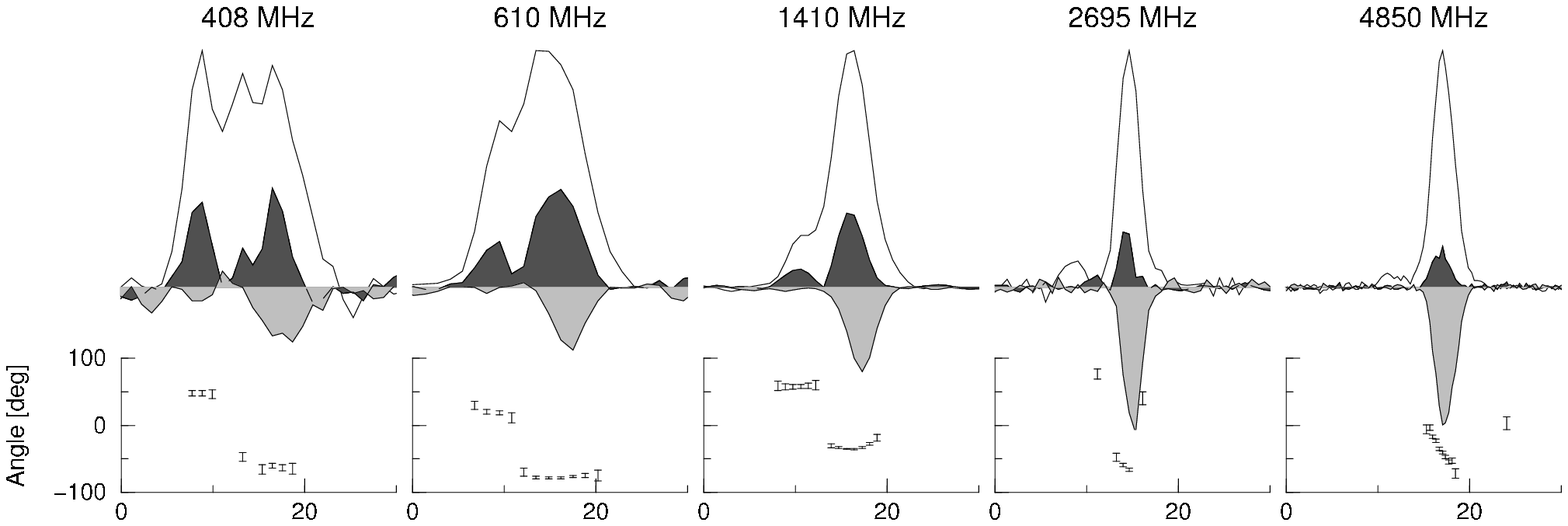}}}
\caption{\em Characteristic polarization behaviour of pulsars. 
The dark-shaded area represents linearly polarized power and 
the light-shaded area circularly polarized power.
  (a): PSR B0355+54. (b): PSR 0525+21.
  (c): PSR B1800-21. (d): PSR 0144+59. From \cite{AvH99}}
\label{Fig:15}
\end{figure}
published \cite{Manchester71a} (Fig. \ref{Fig:13}). Numerous studies 
of the Crab pulsar in polarization have also
 been made \cite{Campbell70}\cite{Graham70}\cite{Manchester72}
\cite{Moffett99}. In 1971 an important result was observed, showing
 that pulsar polarization is constant up to some frequency,
after which it decreases almost linearly \cite{Manchester73}.
This was a result not observed in any other radio 
source and required new interpretation. 

Major surveys of polarisation characteristics of 
larger samples of pulsars started with the access to 
large radio telescopes like the Effelsberg 100-m radio 
telescope \cite{Morris79}\cite{Morris81}\cite{Xilouris91}
\cite{Xilouris95}. The Arecibo telescope was involved in 
polarimetric observations \cite{Rankin89}\cite{Weisberg99}. 
In the southern skies pulsars were studied with the Parkes dish  \cite{Hamilton77}\cite{McCulloch78}\cite{Manchester80}
\cite{Wu93}\cite{Qiao95}\cite{Manchester98}. 
A multi-frequency survey of 300 radio pulsars, at 
frequencies below 1.6 GHz, has 
been conducted \cite{Gould98} with the Lovell telescope 
at Jodrell Bank. The polarization of a large 
sample of pulsars was subsequently studied at the highest 
radio frequencies \cite{AvH97a}\cite{AvH99} up to the 
frequency of 32 GHz (Fig. \ref{Fig:14}). 
Some generalizations about pulsar polarization properties 
can now be made. 

In Fig. \ref{Fig:15} we show the 
polarization evolution with frequency for four characteristic 
types of pulsars. The pulsar B0355+54 
(Fig. \ref{Fig:15}a) begins with low 
linear polarization percentage at low radio frequencies, 
then reaches a maximum (for one component) in the middle 
range of frequencies and finally falls to low polarization 
values at the highest frequency so far observed. The phase 
sweep is linear for the highly polarized component but 
jumps through $90{\hbox{$^{\circ}$}}$ between components. Circular polarization 
is low but evolves in a manner similar to the linear 
polarization. Most pulsars evolve in this manner. 
The second evolution sequence is shown for pulsar 
B0521+21  (Figure \ref{Fig:15}b).  Both components are polarized. 
The degree of polarization falls at high frequencies and the
phase-sweep is S-shaped for both components. In Figure \ref{Fig:15}c 
the polarization-evolution of the pulsar B1800-21 is shown. 
This has the familiar increase and decrease behaviour, 
but in addition, with a considerable circular polarization 
component. A more unusual evolution is seen in Figure 
\ref{Fig:15}d  for the pulsar B0144+59. The linear polarization decreases
 with frequency. However its circular polarization keeps 
increasing up to the highest frequency observed so far. 
   \subsubsection{Millisecond Pulsars (P $<$ 20 ms)}
%
\begin{center}
\begin{figure}[t]
\rotatebox{-90}{\resizebox{9.5cm}{!}{%
\includegraphics[0,0][482,588,0]{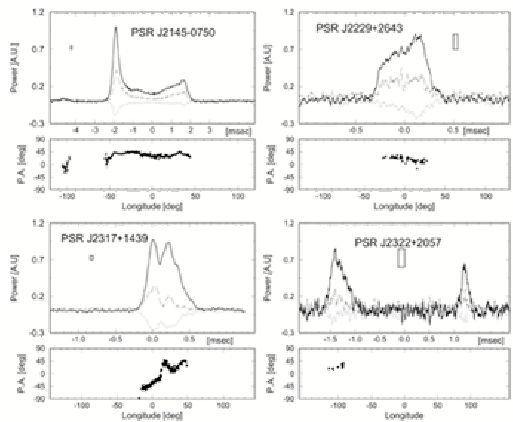}}}
\caption{\em The polarization of millisecond pulsars at 1410 MHz from \cite{Xilouris98}. Four curves are plotted: the total power is shown by
the outer curve in arbitrary units. The linear polarization is shown by the 
dashed curve and the circular polarization by the dotted curve. The linear
polarization position angle is plotted on the lower panel in degrees. All 
curves are plotted against longitude in ms (upper scale) and in degrees
(lower scale)}
\label{Fig:16}
\end{figure}
\end{center}
The discovery of millisecond pulsars \cite{Backer82}
did not lead to immediate studies 
of their polarization. The first reports came in the 90's 
\cite{Thorsett90}\cite{Navarro95}. 
This was due to the fact that millisecond pulsars are, 
on average, an order of magnitude less luminous 
than normal pulsars and thus require very sensitive 
polarimeters. A  major contribution on millisecond 
pulsar polarimetry was published in 1998 \cite{Xilouris98} 
and more recently in 1999 \cite{Stairs99}. The polarization 
characteristics of millisecond pulsars are similar to those of 
 normal pulsars, namely that the linear 
polarization falls to high frequencies \cite{Xilouris98} 
(Fig. \ref{Fig:16}).
   \subsection{Single Pulses}
Most of the polarization observations presented 
so far referred to integrated pulses. 
However it is the polarization of the single pulse that tells us about 
constraints that are necessary for the interpretation 
of their emission mechanism. Early observations
 \cite{Rankin74}\cite{Manchester75b} gave strong indications 
that pulsar radiation is highly polarized. Soon after a strange behaviour
was also detected, e.g that orthogonal 
polarization modes, i.e. abrupt jumps of the position 
angle by {\it $90{\hbox{$^{\circ}$}}$} in consecutive single
pulses are often observed \cite{Manchester75a}\cite{Ganga97}.

%
\begin{figure}[h]
{\resizebox{10.0cm}{8cm}{%
\includegraphics{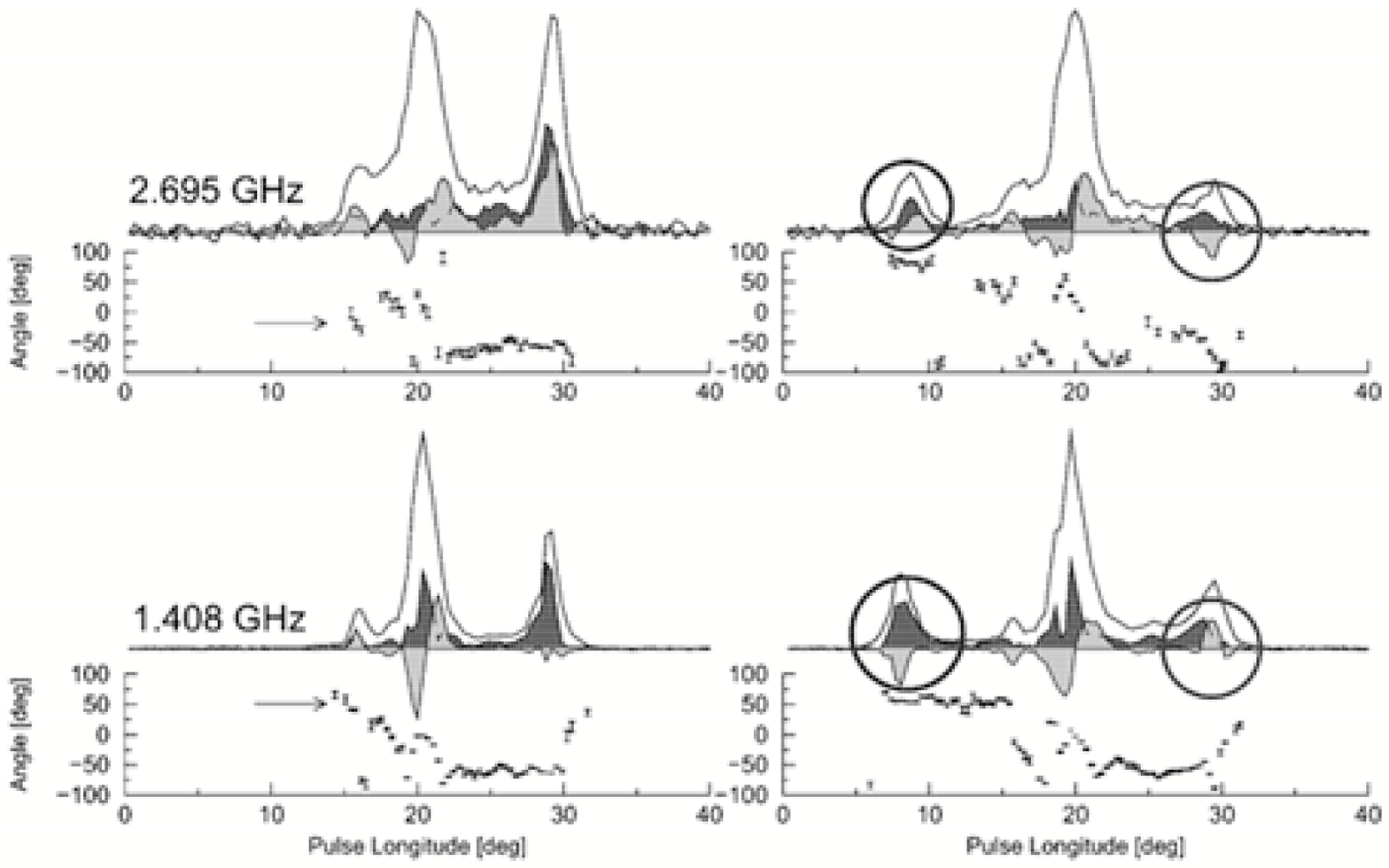}}}
\caption{\em Two examples of individual pulse pairs of PSR B0329+54, 
observed in full polarization. The dark-shaded area represents linearly 
polarized power and the light-shaded area circularly polarized power. 
From \cite{Karastergiou01}}
\label{Fig:17}
\end{figure}
In 1995 a major co-operation project was organized under the 
auspices of the European Pulsar Network. Various telescopes in 
Europe (Effelsberg, Jodrell Bank, Bologna, Westerbork, Torun 
and Pushchino) were used simultaneously to observe pulsars at a 
number of frequencies. Recently the radio telescopes in Ooty and 
GMRT (both in India) have joined this network. Each 
telescope was optimal at some frequency, so that a very wide 
frequency coverage was achieved. Many of the telescopes have 
the capability to observe the full polarization of pulsars as well. 
The results of this major multi-frequency network have shed new 
light on the wide band performance of pulsars as emitters \cite{Karastergiou01}\cite{Karastergiou02}
\cite{Karastergiou03a}\cite{Karastergiou03b}
\cite{Kramer03}.  The fundamental result, that pulsar radio 
emission is basically broad-band, was confirmed.  This is seen in Fig.
 \ref{Fig:8} where single pulses observed at four widely spaced 
frequencies are plotted. Observations of full polarization showed 
even more unusual time sequences. While the total 
intensities correlated rather well, the polarization deviations 
were much higher. In particular the circular polarization (usually 
observed in the conal lobes of a pulse) vary highly from pulse 
to pulse (Fig. \ref{Fig:17}).
   \section{Distributions}
More than 1500 pulsars have been discovered to date, facilitating the statistical 
investigation of their distribution in space, period, period derivative and in
other parameter spaces. These distributions are by now statistically stable and
reliable, not only because of the large number of stars involved in the sample 
but because they cover a large portion ($\sim$1\%) of the pulsar population which 
is believed to be of the order of $10^5$ pulsars in our Galaxy. The figures 
presented below were produced using the data for 1300 pulsars available in 
the recently released ATNF pulsar catalogue 
(www.atnf.csiro.au/research/pulsar/psrcat/) 
\cite{Manchester03}. A few pulsars in the ATNF pulsar catalogue are X-ray
pulsars. Among them there are some with long period. They have been included 
as their properties are very similar to normal radio pulsars. 
   \subsection{Spatial distribution}
%
\begin{figure}[h]
\rotatebox{-90}{\resizebox{6cm}{!}{%
\includegraphics{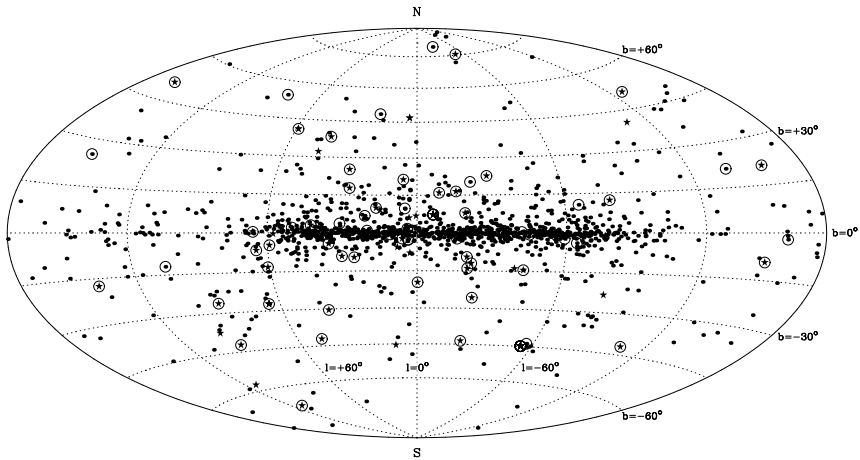}}}
\caption{\em The distribution of 1395 pulsars in
 Galactic Coordinates. Normal pulsars are depicted by dots.
Binary pulsars have a large circle around them. Pulsars 
with p $<$ 20 msec are depicted by stars. Binary millisecond pulsars 
are shown by encircled stars. Figure courtesy of  B. Klein}
\label{Fig:18}
\end{figure}
The distribution of pulsars in galactic coordinates is shown in Fig. \ref{Fig:18} 
in Hammer-Aitoff projection. It is obvious that pulsars are strongly grouped 
along the galactic plane. Millisecond pulsars (many of which are also binary) 
are more isotropically distributed. This effect is due to their inherent weaker 
emission, which allows the detection of, primarily, nearby objects.
   \subsection{Dispersion Measure vs. Galactic Latitude distribution}
%
\begin{figure}[h]
{\resizebox{11.0cm}{!}{%
\includegraphics{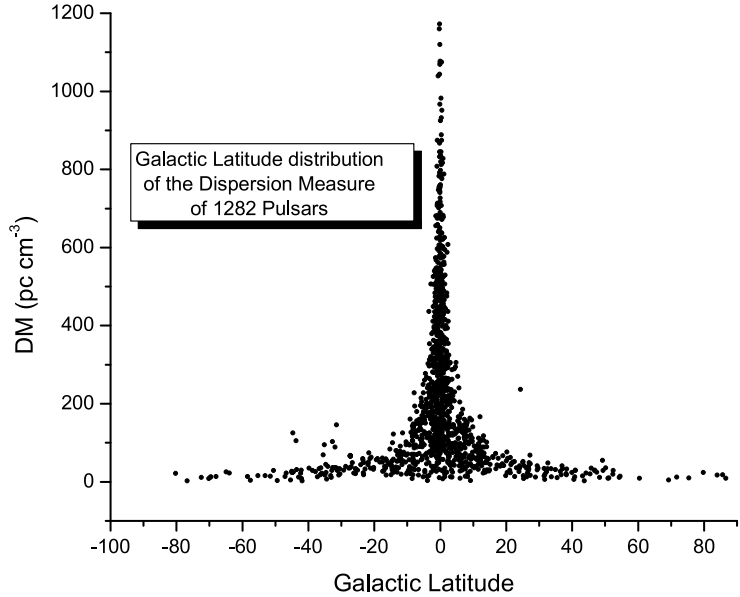}}}
\caption{\em The distribution of the Dispersion Measure 
of 1282 pulsars as a function of their Galactic Latitude}
\label{Fig:19}
\end{figure}
The Dispersion Measure (DM) of pulsars depicts the electron content on the
line of sight, between the star and the observer, $\int{n_e(l)\times dl}$,
where $n_e(l)$ is the electron density and l is the distance. If the average
electron density is known, then the DM is a direct measure of the pulsar
distance. From Fig. \ref{Fig:19} it is obvious (a) that pulsars are tightly 
clustered along the galactic plane (galactic latitude, 
b = 0${\hbox{$^{\circ}$}}$) and (b) that the 
highest DMs are found on this plane. This is due to the 
electron density distribution in the Galaxy, which has been modelled by
several researchers \cite{Lyne85}\cite{Taylor93}. According to 
these models, the electron density peaks at the Galactic centre and falls
off with a scale height of about 70 parsecs away from the plane.

A few pulsars deviate from the smooth 1/e distribution. These objects 
are known to be located behind dense HII regions, e.g the Gum Nebula. 
   \subsection{Period distribution}
%
\begin{figure}[h]
{\resizebox{10.5cm}{!}{%
\includegraphics{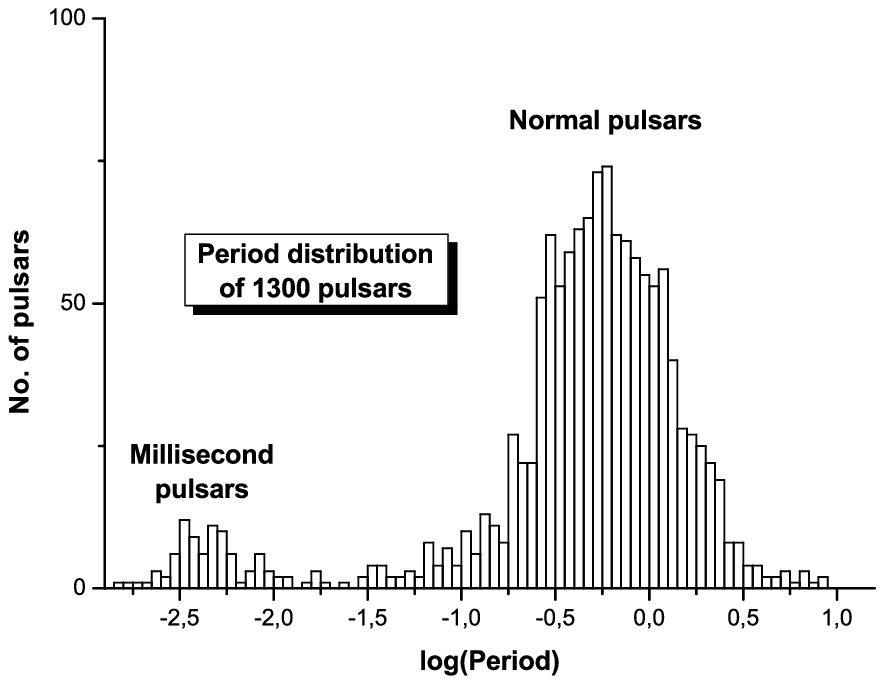}}}
\caption{\em The period distribution of 1300 pulsars}
\label{Fig:20}
\end{figure}
Pulsar periods range from 1.5 ms to about 8.5 s (Fig. \ref{Fig:20}). 
This range of values for
a physical parameter characterizing one species of objects is too wide to 
be explained by a common origin. It is widely accepted that ms pulsars
are recycled pulsars, spun up by accretion processes, during which they
accumulate mass and obtain extra spin (from this mass) from a binary
component. This is confirmed by the data of Fig. \ref{Fig:18}, which
show that most millisecond pulsars are in binary systems.

The histogram of pulsar periods shows a distinct bimodal distribution.
The median of the period distribution of normal pulsars is about 0.65 s,
whereas for millisecond pulsars it is 0.0043 s (4.3 ms). There is a characteristic
lack of pulsars with period around 20 ms.
   \subsection{Period derivative ($\dot P$) distribution}
The $\dot P$ distribution of pulsars (not shown, as it is directly 
connected to Figs. \ref{Fig:20} and \ref{Fig:21}) varies between 
$10^{-21}$ $s s^{-1}$ to $10^{-10}$  $s s^{-1}$. Millisecond pulsars
exhibit a much slower decay (lengthening) of their period. One of the fastest 
decaying period pulsars is the Crab pulsar, whose period slows 
down by 36 ns per day. In general the $\dot P$ 
distribution of pulsars is very similar to the period distribution.

There is a noticeable bimodal distribution. The median of the 
$\dot P$ distribution of normal pulsars is  $2.6 \times 10^{-15}$ 
 $s s^{-1}$, whereas for millisecond pulsars it is of the order of
$2 \times 10^{-20}$  $s s^{-1}$. Five orders of magnitude
slower period decay.
   \subsection{Period -- Period derivative ($\dot P$) distribution}
%
\begin{figure}[h]
\rotatebox{-90}{\resizebox{10.3cm}{!}{%
\includegraphics{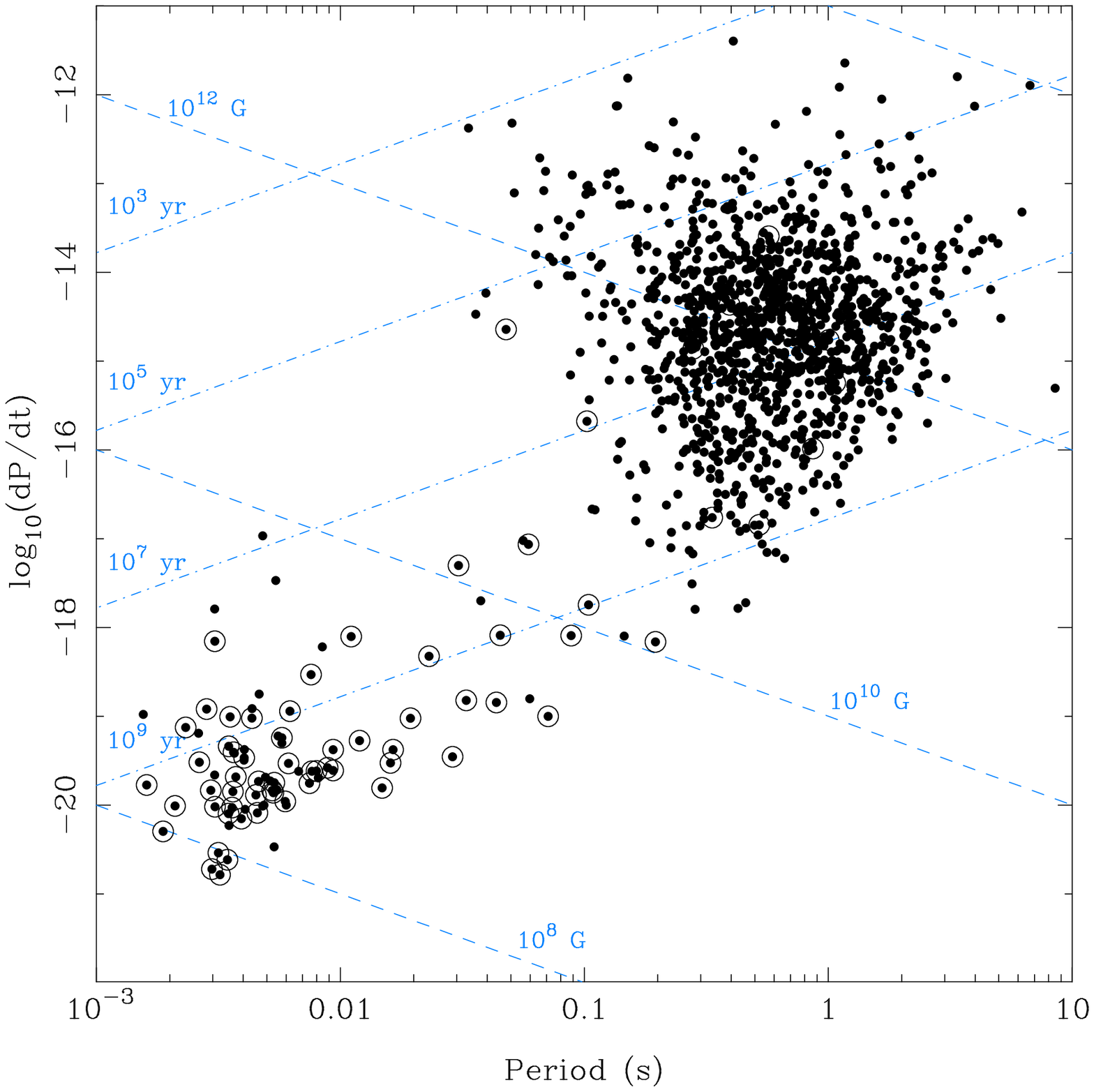}}}
\vspace{1cm}
\caption{\em The P -- $\dot P$ distribution of 1395 pulsars. Binary
pulsars are depicted by encircled dots}
\label{Fig:21}
\end{figure}
\begin{figure}[t]
{\resizebox{11.5cm}{!}{%
\includegraphics{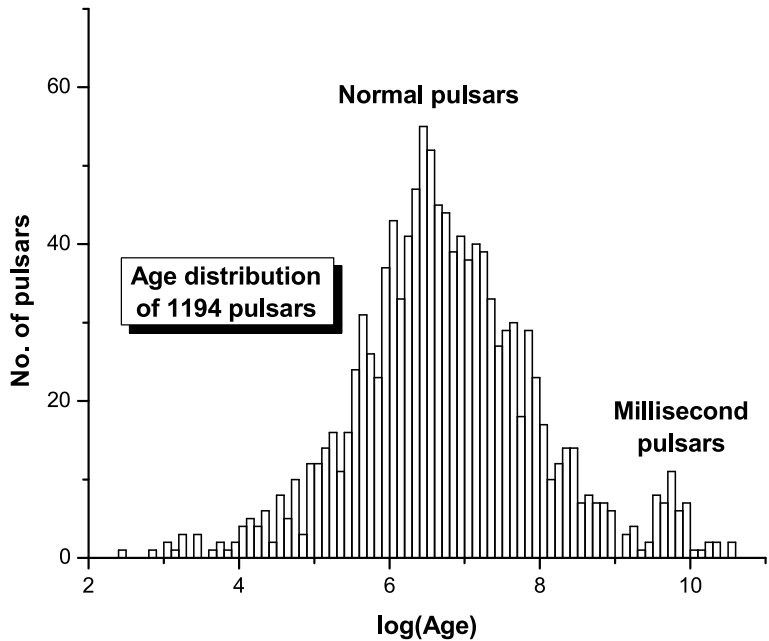}}}
\caption{\em The Age distribution of 1194 pulsars}
\label{Fig:22}
\end{figure}

One of the most important graphical distributions of pulsars is their 
period (P) -- period derivative ($\dot P$) distribution (Fig. \ref{Fig:21}).
As expected from Figures \ref{Fig:20} and \ref{Fig:22} it shows
very clearly the characteristic clustering of normal pulsars (large P, large
$\dot P$) and of millisecond pulsars (small P, small $\dot P$). The vast majority
of normal pulsars are isolated single stars. On the contrary, the majority 
of millisecond pulsars are members of binary star systems. Binary pulsars
located between the two clusters will slowly drift toward the millisecond
cluster in less than $10^8$ years.

Assuming that the decay of pulsar periods is due to their dipole radiation,
their {\it characteristic age} can be calculated from the very simple
expression $ \tau_{char} = {1\over 2} {P\over \dot P}$ years. The ``dash-dot"
lines in Fig. \ref{Fig:21} correspond to lines of constant age. The 
characteristic age of normal pulsars is of the order of $10^8$ years, 
whereas the age of millisecond pulsars is slightly above $10^9$ years. The 
Crab pulsar is the isolated pulsar closest to the $10^3$ years line.

Following classical electrodynamics theory, the surface magnetic field
of pulsars is given by the expression $B_0 = 3.3 \times 10^{19} \times
\sqrt{P \dot P}$ gauss. The ``dash" lines in Fig. \ref{Fig:21} correspond
to lines of constant magnetic field. It is immediately noted that normal
pulsars have a surface magnetic field of about $10^{12}$ gauss, whereas 
the surface magnetic field of millisecond pulsars is much lower, of the order of
$10^9$ gauss.

Finally it should be mentioned that the absence of pulsars in the lower 
right corner of the diagram is due to the existence of a ``death line",
owing to the gradual decaying of the induced electrical potential of pulsars. Slow 
pulsars with low magnetic field cannot develop a large enough potential
above their magnetic poles for discharges (and therefore radiation) to 
take place. The absence of millisecond pulsars below about $10^{10}$ years
indicates that their age cannot exceed the Hubble time (age of the Universe).
%
   \subsection{Age distribution}
The characteristic age ($ \tau_{char} = {1\over 2} {P\over \dot P}$) 
distribution of pulsars (Fig. \ref{Fig:22}) shows the expected bimodal 
distribution attributed to the different ages of normal pulsars and millisecond 
pulsars. The median age of normal pulsars is  $4.7 \times 10^6$ years,
whereas the age of millisecond pulsars is of the order of $5 \times 10^9$ years.
   \subsection{Surface Magnetic Field distribution}
%
\begin{figure}[h]
{\resizebox{11.5cm}{!}{%
\includegraphics{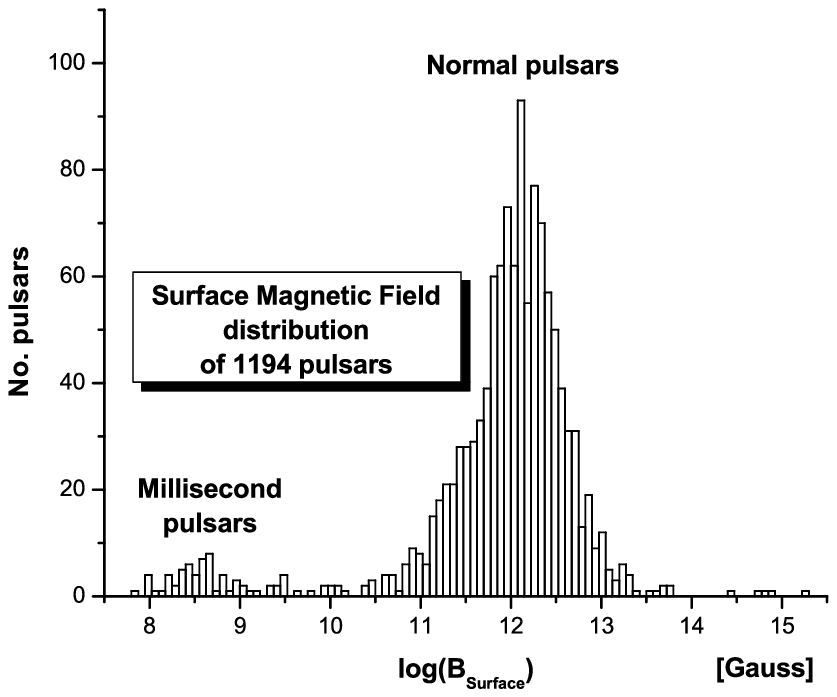}}}
\caption{\em The surface magnetic field distribution of 1194 pulsars}
\label{Fig:23}
\end{figure}
The surface magnetic field ($ B_0 = 3.3 \times 10^{19} \times \sqrt{P \dot P} $) 
distribution of normal pulsars is tightly peaked
at $1.3 \times 10^{12}$ gauss, whereas for millisecond pulsars is much
lower, of the order of $4 \times 10^8$ gauss(Fig. \ref{Fig:23}). In a few stars the surface
magnetic field is larger than $10^{14}$ gauss. This is the surface magnetic 
field of a new class of objects, called {\it magnetars}, first detected
through their X-ray emission \cite{Kouveliotou98}. Magnetars are known
to be neutron stars, which do not always emit at radio wavelengths.
   \subsection{Luminosity distribution}
%
\begin{figure}[h]
{\resizebox{11.5cm}{!}{%
\includegraphics{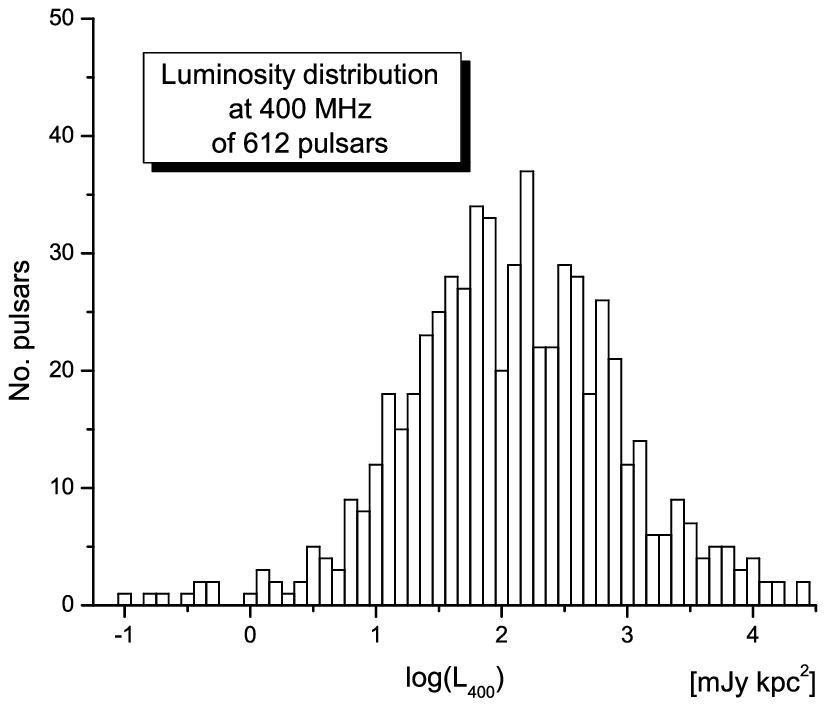}}}
\caption{\em The luminosity distribution at 408 MHz of 612 pulsars}
\label{Fig:24}
\end{figure}
The luminosity distribution at 400 MHz of 612 pulsars is depicted in 
Fig. \ref{Fig:24}. The mean of the gauss fitted distribution is
115 $mJy \ kpc^2$. The luminosity is calculated from the 400 MHz
flux density, assuming that the Dispersion Measure is a true measure
of the distance of each pulsar. This may lead to an over estimate of the
luminosity of pulsars located behind HII regions.
   \subsection{Spectral Index distribution}
%
\begin{figure}[h]
{\resizebox{11.5cm}{!}{%
\includegraphics{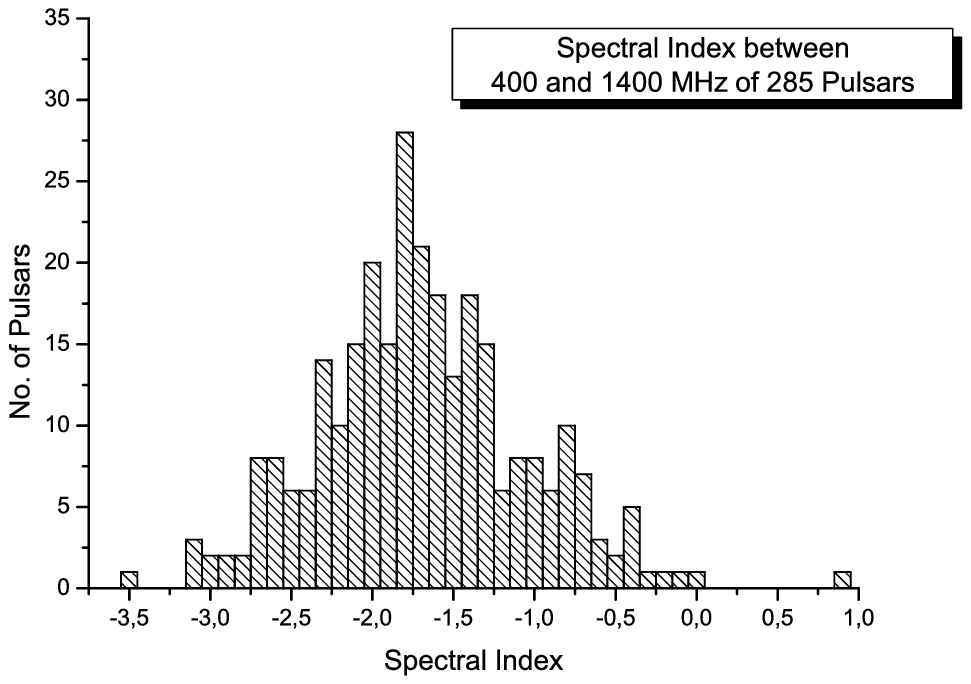}}}
\caption{\em The spectral index distribution of 285 pulsars. Data
  between 400 MHz and 1400 MHz were used}
\label{Fig:25}
\end{figure}
The spectral index distribution between 400 MHz and 1400 MHz 
of 285 pulsars is shown in Fig. \ref{Fig:25}. It is a rather wide
distribution. The bulk of pulsars demonstrate a spectral index
between -3 and 0. The mean of the gauss fitted distribution is
-1.75 $\pm$ 0.1. Pulsars with flat spectral indices are the ones
which should be investigated at high frequencies.
\vspace{0.5cm}
\noindent{\bf Acknowledgements} { JHS acknowledges financial 
support from the Alexander von Humboldt Foundation and the 
Max-Planck-Gesellschaft during his sabbatical from the University 
of Thessaloniki. }. We would like to acknowledge the fact that the
publicly available ATNF pulsar catalogue of 1300 objects has given 
new momentum to pulsar research. We  thank Dr. Axel Jessner for
comments on an early version of this paper. Figures \ref{Fig:18} 
and \ref{Fig:21} were produced by Dr. Bernd Klein (Max-Planck-Institut 
f\"ur Radioastronomie, Bonn). Dr. Michael Kramer provided us with 
unpublished material which was used in some figures. Finally, 
we would like to thank an anonymous referee for useful comments.
%

\end{document}